\documentclass{article}

 \usepackage[main, final]{neurips_2026}

\usepackage[utf8]{inputenc} 
\usepackage[T1]{fontenc}    
\usepackage{hyperref}       
\usepackage{url}            
\usepackage{booktabs}       
\usepackage{amsfonts}       
\usepackage{nicefrac}       
\usepackage{microtype}      
\usepackage{xcolor}         

\usepackage{booktabs}
\usepackage{tabularx}
\usepackage{multirow}
\usepackage{threeparttable}
\usepackage{amsmath}
\usepackage{graphicx}
\usepackage{amsthm}
\usepackage{subcaption}
\usepackage{wrapfig}
\usepackage[most]{tcolorbox}
\usepackage{mdframed}
\usepackage{float}
\usepackage{enumitem}
\usepackage[table]{xcolor}
\definecolor{warmhl}{RGB}{255,224,178}   
\definecolor{warmhdr}{RGB}{255,196,120}  

\usepackage{array}
\usepackage{makecell}
\usepackage{pifont} 
\usepackage{algorithm}
\usepackage{algpseudocode}

\newcommand{\cmark}{\ding{51}} 
\newcommand{\xmark}{\ding{55}}

\title{BiShield-TEE: On the (In-)Security of Unilateral Weight Obfuscation in On-Device TEE-Shielded LLM Partition}

\author{%
  \textbf{Zhiyuan Wang}$^{1,2}$ \quad
  \textbf{Yuezhi Zhou}$^{1,2}$%
  \thanks{Corresponding author.} \quad
  \textbf{Ziqi Zhang}$^{3}$ \\
  \textbf{Zhu Zhang}$^{1}$ \quad
  \textbf{Zhixing Tan}$^{2}$\footnotemark[1] \quad
  \textbf{Yongheng Deng}$^{1}$ \\[0.3em]
  $^{1}$Key Laboratory of Pervasive Computing, Ministry of Education, \\
  Department of Computer Science and Technology, \\
  Tsinghua University, Beijing 100084, China \\
  $^{2}$Zhongguancun Laboratory, Beijing, China \\
  $^{3}$University of Illinois Urbana-Champaign, IL, USA
  \\[0.2em]
  \texttt{\{wzycs24,zhuzhang24\}@mails.tsinghua.edu.cn} \quad
  \texttt{zhouyz@mail.tsinghua.edu.cn} \\
  \texttt{ziqi24@illinois.edu} \quad
  \texttt{zxtan@zgclab.edu.cn} \quad
  \texttt{dengyh1013@gmail.com}
}

\begin{document}

\maketitle

\begin{abstract}
On-device deployment of large language models (LLMs) is increasingly adopted due to latency and privacy requirements, but exposes proprietary models to reverse engineering and intellectual property theft. Trusted Execution Environments (TEEs) provide hardware-enforced isolation for on-device inference, yet most TEE-based parameter obfuscation relies on unilateral matrix transformations that obfuscate only output side.
We identify a vulnerability in this design, termed the \emph{Solution Determinacy Property (SDP)}: the obfuscation secret is uniquely determined, exposing the structure of the true key and enabling attackers to break the protection. 
Under our attacks, representative methods \textsc{GroupCover} and \textsc{ArrowCloak} are reduced to negligible protection, with attack accuracy reaching over $2.4\times$ the full-shielding black-box baseline and recovering near-complete model functionality.
To address this vulnerability, we propose \textsc{BiShield-TEE}, a bilateral framework that induces \emph{Ambiguous Solutions} through joint obfuscation of input and output dimensions, hiding the true key among many equally valid candidates. Experiments across diverse models and datasets show that \textsc{BiShield-TEE} confines attack accuracy to the full-shielding black-box baseline, while achieving higher inference efficiency than prior methods on real hardware.
Our code is available at \url{https://github.com/laduiw/BiShield-TEE}.
\end{abstract}

\section{Introduction}
\label{sec:intro}
The deployment of large language models (LLMs) is increasingly shifting from cloud servers to edge devices, particularly in latency-sensitive scenarios \cite{qin_empirical_2025, wu_first_2024, chen_asynchronous_2025}. While this transition reduces response time and enhances user privacy by enabling local processing \cite{ram_mohan_rao_privacy_2018, mohanty_distillock_2025, lee_occlumency_2019}, it also exposes high-value models to untrusted and potentially adversarial environments. Unlike cloud-based services where models operate as black boxes behind APIs, on-device deployment requires storing and executing models locally, making them vulnerable to white-box attacks \cite{liu2023mirrornet, hou_model_2022, zhang_no_2024}. Given that high-performance models represent valuable intellectual property \cite{yao_survey_2024, DeepIPR, orekondy_knockoff_2019}, protecting them in on-device settings becomes critical \cite{huang_middle_2025, sun_tensorshield_2025}.

Trusted Execution Environments (TEEs) provide hardware-enforced isolation to protect sensitive computations even against privileged adversaries \cite{costan_intel_2016, pinto_demystifying_2019, sev2020strengthening, KV}. However, limited computational resources and memory in edge TEEs \cite{costan_intel_2016, zhang2022teeslice, zhang_groupcover_2024} make full model execution impractical. Existing defenses thus adopt a partitioned execution paradigm \cite{zhang_groupcover_2024, shen_soter_2022, wang_game_2025, sun_shadownet_2023, li_translinkguard_2024, sun_tsqp_2025}, offloading computation-intensive layers to untrusted accelerators while keeping lightweight recovery logic inside the TEE. A common design is to obfuscate each linear layer's weights $\mathbf{W}$ with a secret invertible transformation $\mathbf{T}$ sampled inside the TEE, deploying $\mathbf{W}_{\text{obf}} = \mathbf{W}\mathbf{T}$ on the GPU. At inference, the GPU computes $\mathbf{Y}_{\text{obf}} = \mathbf{X} \mathbf{W}_{\text{obf}}$, and the TEE recovers the true output via $\mathbf{Y} = \mathbf{Y}_{\text{obf}} \mathbf{T}^{-1} = \mathbf{X}\mathbf{W}$. Existing schemes explore various designs of $\mathbf{T}$, seeking to ensure that the obfuscated parameters leak no information about the underlying private weights.

However, current approaches share a common paradigm of \textbf{unilateral obfuscation}, reducing to right-multiplication by a secret $\mathbf{T}$. We identify a fundamental vulnerability of this paradigm, termed \emph{SDP}: The secret $\mathbf{T}$ admits a unique, high-precision estimate $\mathbf{\hat{T}}$ that inherits its structure and enables successful reconstruction, provided the victim is fine-tuned from a public pre-trained model.

Exploiting this structure leakage in practice is non-trivial. Existing attacks~\cite{wang_game_2025} succeed on per-vector obfuscation \cite{shen_soter_2022, sun_shadownet_2023, li_translinkguard_2024, sun_tsqp_2025} through vector direction similarity, where $\hat{\mathbf{T}}$ takes a simple diagonal or permutation form. But these attacks are ineffective against the more elaborate $\hat{\mathbf{T}}$ produced by \textsc{GroupCover} (ICML'24) \cite{zhang_groupcover_2024} and \textsc{ArrowCloak} (USENIX Security'25) \cite{wang_game_2025}. 
We design two dedicated attacks that exploit specific structural patterns in $\hat{\mathbf{T}}$: for \textsc{GroupCover}, $\hat{\mathbf{T}}$ exhibits a block-diagonal structure that we recover through a clustering procedure based on energy-distribution similarity; for \textsc{ArrowCloak}, $\hat{\mathbf{T}}$ admits a low-rank structure that manifests as a spectral anomaly in $\mathbf{W}_\text{obf}$, which we recover via Principal Component Analysis (PCA). Both deliver the first successful attack of these schemes, recovering model functionality nearly identical to the originals.

To prevent the structural leakage exploited by SDP, we propose \textsc{BiShield-TEE}, a bilateral framework that obfuscates both sides of the weight matrix.
We show that this design avoids SDP and induces \emph{Ambiguous Solutions}, yielding an ill-posed system where the true keys are not uniquely recoverable.

We conduct extensive evaluations across diverse models and tasks, including ViT~\citep{dosovitskiy_image_2020}, GPT-2~\citep{radford2019language}, and Llama3.2~\citep{grattafiori_llama_2024}. Our results expose the fragility of existing defenses: attacks on \textsc{GroupCover} and \textsc{ArrowCloak} elevate attack accuracy to $2.44\times$ the black-box baseline, effectively recovering model functionality. By contrast, \textsc{BiShield-TEE} confines attack accuracy to the black-box level. Beyond stronger security, the bilateral design also yields favorable efficiency, running $1.70\times$ and $1.46\times$ faster than \textsc{GroupCover} and \textsc{ArrowCloak}, respectively.

We summarize our contributions as follows:
\begin{itemize}
    \item We identify the \textbf{SDP}, a common vulnerability in unilateral obfuscation, and devise effective attacks on \textsc{GroupCover} and \textsc{ArrowCloak}. By exploiting the structural leakage in each scheme, our attacks recover model functionality close to the white-box baseline.

    \item We propose \textsc{BiShield-TEE}, a bilateral framework that obfuscates input and output sides, providing strong protection while remaining efficient for Transformer architectures.

    \item We empirically show that unilateral defenses degrade to near-white-box security, whereas \textsc{BiShield-TEE} achieves strong protection with improved inference latency on real devices.
\end{itemize}

\section{Related Work}

\subsection{On-device Model Stealing}
Deploying a proprietary model $M_{\text{vic}}$ on edge devices exposes it to model stealing and IP leakage \cite{mohanty_distillock_2025, oliynyk2023know}. We assume the adversary can access all information outside the TEE. With white-box access to unprotected parameters, the adversary directly extracts weights to build a high-fidelity surrogate $M_{\text{sur}}$ \cite{wang_game_2025}, whereas under full-shielding black-box access, $M_{\text{vic}}$ is fully protected inside the TEE and the adversary has no information about its parameters \cite{hanzlik_mlcapsule_2021}. Our defense aims to reduce adversarial capability from white-box to black-box, limiting the effectiveness of $M_{\text{sur}}$.

\subsection{Trusted Execution Environment}
TEE is a secure, hardware-enforced area within a processor that protects code and data integrity and confidentiality from the host OS \cite{TEEWhat}.
A prominent implementation, Intel Software Guard Extensions (SGX)~\cite{costan_intel_2016}, provides a private memory region called an enclave in which contents are encrypted by the CPU and isolated by hardware from external access.
Crucially, SGX supports remote attestation \cite{zhang_attestllm_2025}, enabling a remote party to verify the inference code's integrity before provisioning proprietary model weights. Following prior work \cite{zhang_no_2024, sun_tensorshield_2025, zhang_groupcover_2024, shen_soter_2022, wang_game_2025, sun_shadownet_2023}, \textsc{BiShield-TEE} adopts the TEE as the root of trust to securely store obfuscation keys and execute lightweight de-obfuscation, while offloading the bulk of computation to untrusted accelerators.


\begin{table}[t]
\caption{Comparative analysis of TEE-based weight obfuscation schemes.}
\label{tab:compare_methods}
\small
\setlength{\tabcolsep}{5pt}
\renewcommand{\arraystretch}{1.2}
\begin{tabular}{@{}llllc@{}}
\toprule
\textbf{Paradigm} & \textbf{Method} & \textbf{Venue} & \textbf{Mechanism} & \textbf{SDP-free} \\
\midrule
\multirow{4}{*}{\shortstack[l]{Unilateral\\Per-vector}}
  & SOTER~\cite{shen_soter_2022}        & ATC'22        & Per-vector scaling               & \xmark \\
  & ShadowNet~\cite{sun_shadownet_2023} & S\&P'23       & Per-vector scaling + permutation & \xmark \\
  & TLG~\cite{li_translinkguard_2024}   & MM'24         & Per-vector permutation               & \xmark \\
  & TSQP~\cite{sun_tsqp_2025}           & S\&P'25       & Quantized per-vector scaling     & \xmark \\
\midrule
Unilateral mul.   
  & \textsc{GroupCover}~\cite{zhang_groupcover_2024} & ICML'24      & Group columns, mix within group & \xmark \\
\midrule
Unilateral add.    
  & \textsc{ArrowCloak}~\cite{wang_game_2025}        & USENIX Sec'25 & Add shared mask to all columns  & \xmark \\
\midrule
\textbf{Bilateral} (ours)
  & \textsc{BiShield-TEE} & this work & Joint input-output obfuscation & \cmark \\
\bottomrule
\end{tabular}
\vspace{-6pt}
\end{table}

\subsection{Partitioned Execution in TEE-Assisted Inference}
Since executing modern LLMs entirely inside TEEs is impractical \cite{kaplan_scaling_2020, coppolino_experimental_2025}, prior work adopts a partitioned paradigm \cite{zhang_no_2024, zhang_groupcover_2024, shen_soter_2022, wang_game_2025, sun_shadownet_2023} that offloads heavy matrix multiplications to untrusted GPUs while keeping key management inside the TEE. A common design obfuscates weights by right-multiplication with a secret $\mathbf{T}$ acting only on the output side of parameter $\mathbf{W}$, with $\mathbf{T}$ typically given a special structure so that $\mathbf{T}^{-1}$ is cheap to apply inside the TEE.

These unilateral schemes differ in how they instantiate $\mathbf{T}$, and Table~\ref{tab:compare_methods} groups them into three types. The first applies per-vector obfuscation such as random scaling or permutation, which is insecure under fine-tuning from public pre-trained checkpoints \cite{zhang_no_2024, zhang2022teeslice, wang_game_2025, xiong2025loro}. The second uses unilateral multiplicative obfuscation (e.g., \textsc{GroupCover} \cite{zhang_groupcover_2024}), mixing output features within groups. The third uses unilateral additive obfuscation (e.g., \textsc{ArrowCloak} \cite{wang_game_2025}), adding a scaled shared mask to each column. All three remain unilateral, confining obfuscation to the output side of $\mathbf{W}$. 
\textsc{Amulet}~\cite{mao2025amuletfastteeshieldedinference} also employs bilateral weight obfuscation, with its design centered on reducing TEE interactions during inference. Our work focuses on the SDP vulnerability of unilateral obfuscation and develops lightweight bilateral transformations to resist reconstruction attacks.
For intermediate features, we follow prior work \cite{tramer_slalom_2019, sun_shadownet_2023} and use integrity verification and one-time pad (OTP) protection~\cite{tramer_slalom_2019}, with implementation details provided in Appendix~\ref{app:background}.

\section{Threat Model}
\textbf{Scenario.} Following prior TEE-based on-device inference settings~\cite{zhang_no_2024, zhang_groupcover_2024, wang_game_2025, li_translinkguard_2024, sun_tsqp_2025}, we consider a model owner deploying a proprietary model \(M_{\mathit{vic}}\) to user-controlled devices with TEEs, where an obfuscated model \(M_{\mathit{obf}}\) runs outside the TEE on the GPU while secrets and recovery logic remain inside TEE. The defender aims to reduce the adversary's effective capability from white-box to black-box, ensuring any extracted surrogate \(M_{\mathit{sur}}\) has substantially lower utility than \(M_{\mathit{vic}}\). The attacker is a malicious device user with full control over \(M_{\mathit{obf}}\), runtime traces, and the software stack outside the TEE, and knows the defense mechanism but cannot access TEE-internal secrets. 

\textbf{Prior Knowledge.} Following prior work and practical deployment settings \cite{zhang_no_2024, zhang_groupcover_2024, wang_game_2025, sun_tsqp_2025}, we assume that $M_{\mathit{vic}}$ is fine-tuned from a publicly available pre-trained model $M_{\mathit{pre}}$. The parameter reconstruction stage of our attack is query-free, relying solely on $W_{\mathit{pre}}$ and the obfuscated weights $W_{\mathit{obf}}$. Queries to $M_{\mathit{vic}}$ are required only for downstream task adaptation (e.g., retraining the surrogate's classification head), and the query budget is capped at less than 1\% of the training data size, consistent with prior work \cite{zhang_no_2024, sun_tensorshield_2025, zhang_groupcover_2024, wang_game_2025, li_translinkguard_2024}.

\section{Attacks on Unilateral Defenses}
\label{sec:attacks}

In this Section, we formalize SDP and instantiate concrete attacks on unilateral defenses.

\textbf{Notation.}
We denote the weight matrices of the victim, obfuscated, and pre-trained models by $\mathbf{W}_{\text{vic}}, \mathbf{W}_{\text{obf}}$, and $\mathbf{W}_{\text{pre}}$, respectively. The $i$-th column is denoted by $\mathbf{w}^{(i)}$. Weight matrices are $\mathbf{W} \in \mathbb{R}^{d \times n}$ with input/output dimensions $d$ and $n$, and features $\mathbf{X} \in \mathbb{R}^{l \times d}$ with sequence length $l$.

\subsection{Solution Determinacy Property in Unilateral Obfuscation}
\label{subsec:sdp}


\begin{wrapfigure}{r}{0.52\linewidth}
  \centering
  \vspace{-\baselineskip}
  \includegraphics[width=\linewidth]{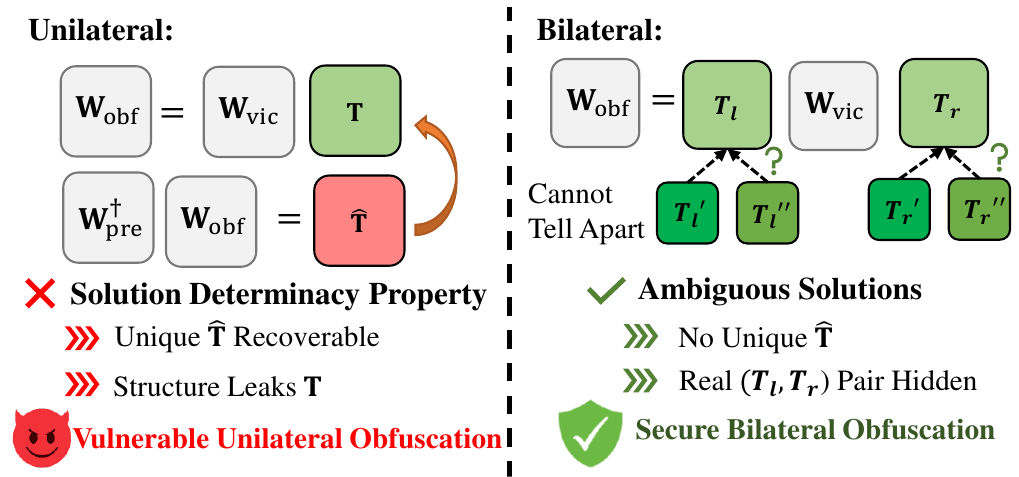}
  \caption{Comparison of unilateral and bilateral obfuscation.}
  \label{fig:uni_bi}
  \vspace{-\baselineskip}
\end{wrapfigure}

Under partitioned TEE inference, outsourced computation is dominated by linear matrix multiplications \cite{ zhang_groupcover_2024, shen_soter_2022,wang_game_2025, sun_shadownet_2023}. To preserve correctness, unilateral defenses apply a right-side transformation $\mathbf{T} \in \mathbb{R}^{n \times n}$ to obtain $\mathbf{W}_{\text{obf}} = \mathbf{W}_{\text{vic}} \mathbf{T}$ (Figure~\ref{fig:uni_bi}). Since the victim is fine-tuned from a public checkpoint, we can write $\mathbf{W}_{\text{vic}} = \mathbf{W}_{\text{pre}} + \boldsymbol{\Delta}$ with a small fine-tuning increment $\boldsymbol{\Delta}$. The adversary can therefore approximately solve for $\mathbf{T}$ from the public $\mathbf{W}_{\text{pre}}$ and the observed $\mathbf{W}_{\text{obf}}$ via least-squares:
\begin{equation}
\label{T_hat}
\hat{\mathbf{T}} = \arg\min_{\mathbf{T}} \| \mathbf{W}_{\text{obf}} - \mathbf{W}_{\text{pre}} \mathbf{T} \|_F^2 = \mathbf{W}_{\text{pre}}^{\dagger} \mathbf{W}_{\text{obf}} \approx (\mathbf{I} + \mathbf{W}_{\text{pre}}^{\dagger} \boldsymbol{\Delta}) \mathbf{T},
\end{equation}
where $\mathbf{W}_{\text{pre}}^{\dagger}$ denotes the Moore--Penrose pseudoinverse. Since $\boldsymbol{\Delta}$ is small, $\hat{\mathbf{T}}$ stays close to $\mathbf{T}$ and preserves its structural properties. A dense random $\mathbf{T}$ would leave $\hat{\mathbf{T}}$ devoid of exploitable structure, but is impractical inside TEEs since applying $\mathbf{T}^{-1}$ during recovery becomes prohibitively expensive. Existing defenses therefore impose structural constraints on $\mathbf{T}$ (e.g., sparsity or low-rank structure), which are inherited by $\hat{\mathbf{T}}$ and allow the adversary to recover the true transformation from its structure. We refer to this inherent vulnerability as SDP.


SDP directly explains why the per-vector schemes \cite{shen_soter_2022, sun_shadownet_2023, li_translinkguard_2024, sun_tsqp_2025} in Table~\ref{tab:compare_methods} are insecure: their permutation- or scaling-based obfuscation causes $\hat{\mathbf{T}}$ to concentrate at the true permutation or scaling locations, so the key can be read off by selecting the maximum in each row or column. Prior attacks~\cite{wang_game_2025} also succeed through vector direction similarity, but cannot extend to \textsc{GroupCover} or \textsc{ArrowCloak}, whose designs further obfuscate the column directions and eliminate the direction similarity.

\begin{figure*}[htbp]
  \centering
  \includegraphics[width=\textwidth]{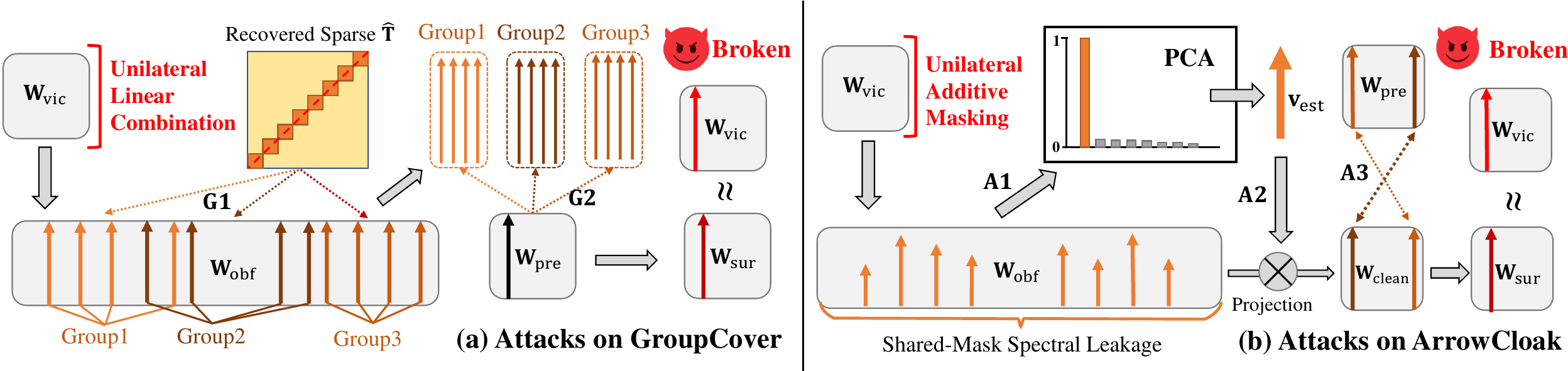}
    \caption{Attacks against unilateral model protection schemes.
    \textbf{(a) Attack on \textsc{GroupCover}.}
    Step G1 recovers the block-diagonal grouping from $\hat{\mathbf{T}}$. Step G2 reconstructs $\mathbf{W}_\text{vic}$ within each group.
    \textbf{(b) Attack on \textsc{ArrowCloak}.}
    Step A1 extracts the shared mask direction via PCA. Step A2 denoises $\mathbf{W}_\text{obf}$ via orthogonal projection. Step A3 restores $\mathbf{W}_\text{vic}$ via least-squares reconstruction.
    }
    \label{fig:pic_attack}
    \vspace{-6pt}
\end{figure*}

We close this gap by exploiting the structures of $\hat{\mathbf{T}}$ to attack \textsc{GroupCover} and \textsc{ArrowCloak}. Since these structures take different forms in each scheme, we design two dedicated attacks, illustrated in Figure~\ref{fig:pic_attack}, that reconstruct the protected weights. The original defense details of \textsc{GroupCover} and \textsc{ArrowCloak} are deferred to Appendix~\ref{app:methods}.

\subsection{Breaking Unilateral Multiplicative Schemes: \textsc{GroupCover} Case Study}
\label{subsec:attack_gc}

\textsc{GroupCover} \cite{zhang_groupcover_2024} partitions the output columns into $n/s$ groups of size $s$, applies an independent random $s \times s$ linear mixing within each group, and permutes columns to hide group membership, yielding a block-diagonal transformation under random row and column permutations:
\begin{equation}
\label{T_gc}
\mathbf{T}_{\textsc{GroupCover}} = \mathbf{\Pi}_1 \, \mathrm{blockdiag}(\mathbf{T}_1, \dots, \mathbf{T}_{n/s}) \, \mathbf{\Pi}_2, \quad \mathbf{T}_i \in \mathbb{R}^{s \times s}.
\end{equation}
The least-squares estimate $\hat{\mathbf{T}}$ inherits this block-sparse structure: each column concentrates its mass on a small subset of $s$ indices, and the group size $s$ can be inferred from the number of dominant entries per row or column. We exploit this leakage via a two-stage attack shown in Figure~\ref{fig:pic_attack}(a).


\textbf{G1: Group recovery.}\
Columns belonging to the same group share a common set of $s$ dominant indices in $\hat{\mathbf{T}}$, while entries outside the group are close to zero. Starting from each unassigned column, we identify the $s-1$ columns with the greatest overlap in the row indices of their largest-magnitude entries and assign them to the same group. Repeating this process until all columns are assigned recovers the grouping $\{G_1, \dots, G_{n/s}\}$.

\textbf{G2: Parameter reconstruction.}
Given the recovered groups, the global reconstruction problem decomposes into independent low-dimensional subproblems. For each column of $\mathbf{W}_{\text{pre}}$, we identify candidate groups in $\mathbf{W}_{\text{obf}}$ from the row indices of its largest-magnitude entries, perform local least-squares reconstruction within each group, and select the result with the highest cosine similarity.
As formally proved in Appendix~\ref{groupcover_proof}, the parameter recovery error is only an $s/d$ fraction of that in the black-box setting; with the default $s=4$ and matrix dimension $d$, we have $s \ll d$, leading to highly accurate recovery.


\subsection{Breaking Unilateral Additive Scheme: \textsc{ArrowCloak} Case Study}
\label{subsec:attack_ac}

\textsc{ArrowCloak} \cite{wang_game_2025} applies per-column scaling controlled by $\mathbf{p}$ and a column permutation $\mathbf{\Pi}$, together with an additive mask shared across all columns, where the direction of the mask is determined by a secret coefficient vector $\mathbf{k}$ and its per-column weights are given by $\mathbf{q}$:
\begin{equation}
\label{T_arrow}
\mathbf{T}_{\textsc{ArrowCloak}} = \bigl(\operatorname{diag}(\mathbf{p}) + \mathbf{k}\mathbf{q}^{\top}\bigr)\mathbf{\Pi}, \quad \mathbf{p}, \mathbf{k}, \mathbf{q} \in \mathbb{R}^n.
\end{equation}
This transformation injects a shared vector $\mathbf{v} := \mathbf{W}_{\text{vic}}\mathbf{k} \in \mathbb{R}^d$ into every column of $\mathbf{W}_{\text{obf}}$, scaled by a per-column factor. The resulting low-rank structure exposes $\mathbf{v}$ as a dominant direction in $\mathbf{W}_{\text{obf}}$, which we exploit via a three-step attack as shown in Figure~\ref{fig:pic_attack}(b).

\textbf{A1: Mask extraction.}
Since the shared mask dominates the variance of $\mathbf{W}_{\text{obf}}$, we estimate its direction via PCA:
\begin{equation}
\mathbf{v}_{\text{est}} \approx \mathrm{PCA}_1(\mathbf{W}_{\text{obf}}).
\end{equation}

\textbf{A2: Denoised alignment.}
The additive term distorts column-wise similarity, making direct matching unreliable. We remove this component by projecting onto the orthogonal complement of $\mathbf{v}_{\text{est}}$:
\begin{equation}
\mathbf{W}_{\text{clean}} = \mathbf{W}_{\text{obf}} - \mathbf{v}_{\text{est}} (\mathbf{v}_{\text{est}}^\top \mathbf{W}_{\text{obf}}),
\end{equation}
and recover the column permutation by computing cosine similarities between normalized columns of $\mathbf{W}_{\text{clean}}$ and $\mathbf{W}_{\text{pre}}$.

\textbf{A3: Parameter reconstruction.}
With column alignment established, each column is reconstructed by solving a least-squares problem over the obfuscated weights and the estimated mask:
\begin{equation}
\mathbf{w}_{\text{sur}}^{(i)} = \hat{\alpha}_i \mathbf{w}_{\text{obf}}^{(i)} + \hat{\beta}_i \mathbf{v}_{\text{est}}.
\end{equation}
Appendix~\ref{app:attack} provides additional details of our attacks against \textsc{GroupCover} and \textsc{ArrowCloak}, including pseudocode, diagnostic visualizations and complexity analysis.

\section{Design of \textsc{BiShield-TEE}}
\label{sec:method}
Since the above attacks all exploit SDP, \textsc{BiShield-TEE} breaks it by applying bilateral obfuscation over both row and column spaces. For a weight matrix \(\mathbf{W}_{\text{vic}} \in \mathbb{R}^{d \times n}\), the protected form is
\[
\mathbf{W}_{\text{obf}}
= \mathbf{T}_l\, \mathbf{W}_{\text{vic}}\, \mathbf{T}_r,
\quad
 \mathbf{T}_l \in \mathbb{R}^{d \times d},\;
\mathbf{T}_r \in \mathbb{R}^{n \times n},
\]
where \(\mathbf{T}_l = \mathbf{H}_l \mathbf{D}_l\) and \(\mathbf{T}_r = \mathbf{D}_r \mathbf{H}_r\).
Each \(\mathbf{D}\) is a diagonal scaling matrix that perturbs the magnitude of every dimension, and each \(\mathbf{H}\) is a Householder matrix \cite{householder1958unitary},
\[
\mathbf{H} = \mathbf{I} - 2\mathbf{v}\mathbf{v}^\top, \; \|\mathbf{v}\|=1,
\]
that perturbs directional structure.
Since Householder matrices are involutory ($\mathbf{H}^{-1}=\mathbf{H}$), the inverses simplify to $\mathbf{T}_l^{-1} = \mathbf{D}_l^{-1}\mathbf{H}_l$ and $\mathbf{T}_r^{-1} = \mathbf{H}_r\mathbf{D}_r^{-1}$, which can be computed efficiently inside the TEE. We then demonstrate this framework on linear layers and self-attention.

\subsection{Linear Layers}
By applying the inverse transformation to the right-side of the input \(\mathbf{X} \in \mathbb{R}^{l \times d}\), the left-side obfuscation of \(\mathbf{W}\) is implicitly cancelled. During inference, \(\mathbf{X}\) is preprocessed inside the TEE as
\[
\mathbf{X}' = \mathbf{T}_m\, \mathbf{X}\, \mathbf{T}_l^{-1}
= \mathbf{T}_m\, \mathbf{X}\, \mathbf{D}_l^{-1} \mathbf{H}_l.
\]
where \(\mathbf{T}_m = \mathbf{H}_m \mathbf{D}_m\) is sampled per forward pass inside the TEE as an independent scale-plus-Householder transformation that protects intermediate activations.
GPU execution then yields
\[
\mathbf{Y}'
= \mathbf{X}' \mathbf{W}_{\text{obf}}
= \mathbf{T}_m\, \mathbf{X}\,
(\mathbf{D}_l^{-1} \mathbf{H}_l \mathbf{H}_l \mathbf{D}_l)\,
\mathbf{W}_{\text{vic}}\, \mathbf{T}_r
= \mathbf{T}_m\, \mathbf{Y}\, \mathbf{T}_r.
\]
The TEE recovers the output via \(\mathbf{Y} = \mathbf{T}_m^{-1}\, \mathbf{Y}'\, \mathbf{T}_r^{-1}\).

\subsection{Self-Attention}

\begin{figure}[htbp]
  \centering
  \includegraphics[width=\linewidth]{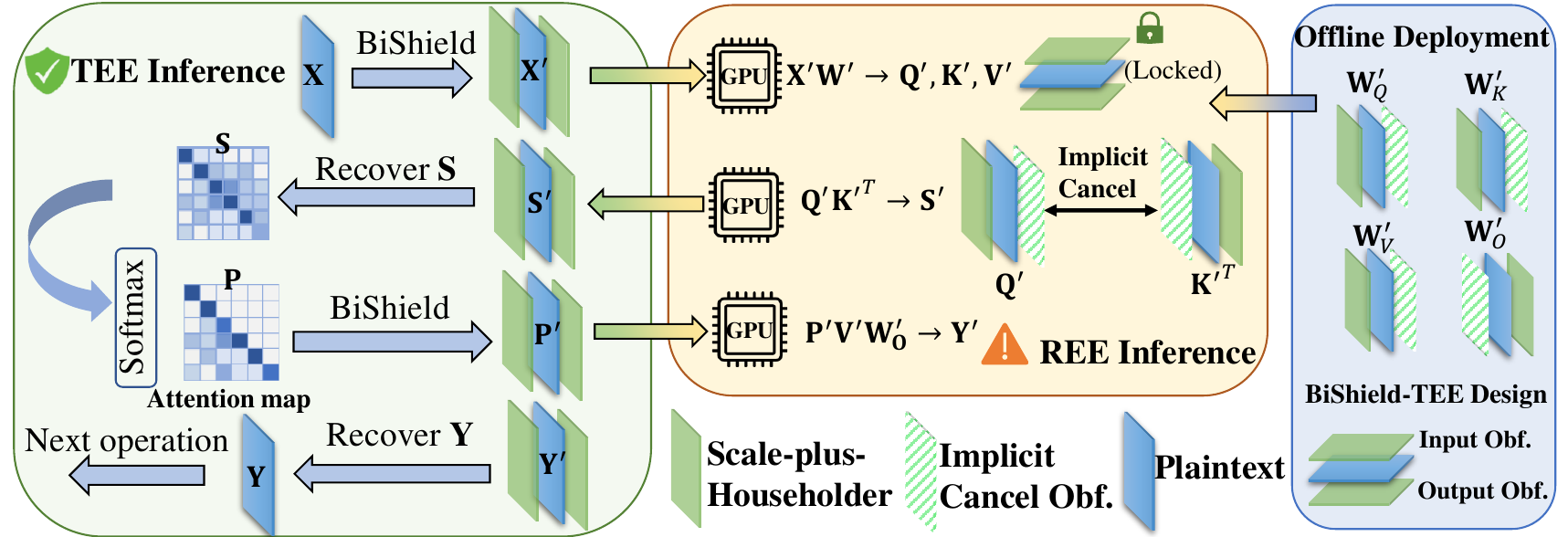}
  \caption{Overview of \textsc{BiShield-TEE} for self-attention layers.}
  \label{fig:overview}
\end{figure}

Figure~\ref{fig:overview} illustrates the \textsc{BiShield-TEE} for self-attention.
Standard self-attention~\cite{vaswani_attention_2017} computes
\[
\mathbf{Q} = \mathbf{X}\mathbf{W}_Q,\quad
\mathbf{K} = \mathbf{X}\mathbf{W}_K,\quad
\mathbf{V} = \mathbf{X}\mathbf{W}_V,
\]
\[
\mathbf{S} = \mathbf{Q}\mathbf{K}^{\top},\quad
\mathbf{P} = \mathrm{softmax}\!\left(\tfrac{\mathbf{S}}{\sqrt{d}}\right),\quad
\mathbf{Y} = \mathbf{P}\mathbf{V}\mathbf{W}_O.
\]
Each projection matrix is obfuscated bilaterally as 
\(\mathbf{W}_Q' = \mathbf{T}_Q^{l}\mathbf{W}_Q\mathbf{T}_Q^{r}\).
With runtime preprocessing \(\mathbf{X}' = \mathbf{T}_Q\mathbf{X}(\mathbf{T}_Q^{l})^{-1}\), we obtain
\[
\mathbf{Q}' = \mathbf{T}_Q\mathbf{Q}\mathbf{T}_Q^{r}.
\]
 Similarly, the transformed projections for keys and values are \(\mathbf{K}' = \mathbf{T}_K\mathbf{K}\mathbf{T}_K^{r}\) and
\(\mathbf{V}' = \mathbf{T}_V\mathbf{V}\mathbf{T}_V^{r}\).
By coupling the two right-side keys via $(\mathbf{T}_K^{r})^{\top} = (\mathbf{T}_Q^{r})^{-1}$, they cancel implicitly in the dot product:
\[
\mathbf{S}' = \mathbf{Q}'\mathbf{K}'^{\top}
= \mathbf{T}_Q\,\mathbf{Q}\,
\mathbf{K}^{\top}\mathbf{T}_K^{\top}
= \mathbf{T}_Q\,\mathbf{S}\,\mathbf{T}_K^{\top}.
\]
The TEE restores \(\mathbf{S} = \mathbf{T}_Q^{-1}\mathbf{S}'(\mathbf{T}_K^{\top})^{-1}\), applies softmax to obtain \(\mathbf{P}\), and re-obfuscates as \(\mathbf{P}' = \mathbf{T}_P\,\mathbf{P}\,\mathbf{T}_V ^{-1}\).
With \(\mathbf{W}_O' = (\mathbf{T}_V^{r})^{-1}\mathbf{W}_O\,\mathbf{T}_O\), the \(\mathbf{T}_V^{r}\) factors cancel implicitly:
\[
\mathbf{Y}'
= \mathbf{P}'\,\mathbf{V}'\,\mathbf{W}_O'
= \mathbf{T}_P\,\mathbf{Y}\,\mathbf{T}_O.
\]
and the TEE recovers \(\mathbf{Y} = \mathbf{T}_P^{-1}\mathbf{Y}'\mathbf{T}_O^{-1}\).

\subsection{Efficiency and Security Analysis}
TEE-side computation is limited to Householder and scale transformations, while dense matrix multiplications are offloaded to GPUs. Since applying a Householder matrix reduces to two matrix-vector products, the recovery overhead remains low. The number of TEE transitions is also unchanged from unilateral schemes~\cite{zhang_groupcover_2024,wang_game_2025}, since both unilateral and bilateral schemes share the same workflow: obfuscation is applied before offloading inputs and reversed once results return to the TEE.


From a security perspective, bilateral obfuscation turns key recovery into an ill-posed problem: given $\mathbf{W}_{\text{obf}} = \mathbf{T}_l \mathbf{W}_{\text{vic}} \mathbf{T}_r$ and a pre-trained model, a continuum of scale-plus-Householder pairs remain observationally indistinguishable up to the fine-tuning slack. This set of \textbf{Ambiguous Solutions} conceals the genuine keys among many equally valid candidates. Whereas unilateral defenses allow the adversary to readily obtain a structurally constrained approximation of the true key, bilateral obfuscation makes such recovery substantially harder. A formal analysis is provided in Appendix~\ref{app:security}.

\section{Experiments}

To evaluate the effectiveness of our approach, we focus on the following key questions:

\begin{tcolorbox}[
    colback=gray!10,
    colframe=black,
    arc=4pt,
    boxrule=1pt,
    left=8pt,
    right=8pt,
    top=6pt,
    bottom=6pt
]
\begin{itemize}[leftmargin=1.5em, itemsep=4pt, topsep=0pt, parsep=0pt]
    \item \textbf{RQ1:} Are representative unilateral defenses secure against our reconstruction attacks?
    \item \textbf{RQ2:} Can bilateral design withstand the same or more adaptive attacks?
    \item \textbf{RQ3:} How efficient is bilateral protection on real-world hardware?
\end{itemize}
\vspace{-5pt}
\end{tcolorbox}

\subsection{Experimental Setup}

\textbf{Models and Datasets.}
We evaluate on five representative architectures spanning vision and language tasks: ViT-Base, ViT-Large, GPT2-Base, Llama3.2-1B and Llama3.2-3B. For vision tasks, ViT models are evaluated on CIFAR-10, CIFAR-100~\cite{krizhevsky2009learning}, and Food-101~\cite{bossard2014food}. For NLP, GPT2-Base is evaluated on four GLUE~\cite{wang2018glue} tasks, and Llama3.2-1B and Llama3.2-3B are evaluated on both GSM8K~\cite{cobbe2021gsm8k} and IFEval~\cite{zhou2023ifeval} to assess mathematical reasoning and instruction-following capabilities.

\textbf{Baselines and Settings.}
The white-box baseline with full plaintext access and the black-box baseline with pre-trained weight substitution represent the upper and lower bounds of attacker capability, respectively. To enable comparison across tasks with different baseline difficulties, we additionally report \textbf{Rel. Black}, the ratio of attack accuracy to the black-box baseline. Values close to $1\times$ indicate that the adversary is reduced to the black-box level, while large values indicate that protection has effectively failed. 
For Llama3.2, surrogate models are evaluated with lm-evaluation-harness~\cite{eval-harness}.
We focus on \textsc{GroupCover} and \textsc{ArrowCloak} as representative unilateral schemes, and additionally include \textsc{TLG}~\cite{li_translinkguard_2024} as a per-vector obfuscation scheme in Table~\ref{tab:compare_methods}, though this category has already been broken by prior work~\cite{wang_game_2025}.
Further details on the experimental pipeline, hyperparameters, and implementation are provided in Appendix~\ref{appendix:setup}.

\begin{table}[htbp]
\centering
\scriptsize
\setlength{\tabcolsep}{4pt}
\renewcommand{\arraystretch}{1.05}
\caption{Attack performance against unilateral defenses. We report our attack accuracy and its ratio to the black-box baseline (Rel. Black). Higher values indicate failed protection (closer to no-shield white-box); lower values indicate effective defense (closer to full-shielding black-box).}
\label{tab:attack_exp}
\resizebox{\textwidth}{!}{%
\begin{tabular}{llccccccccc}
\toprule
\multirow{2}{*}{\textbf{Model}} & \multirow{2}{*}{\textbf{Dataset}} & \multicolumn{2}{c}{\textbf{\textsc{TLG}}} & \multicolumn{2}{c}{\textbf{\textsc{GroupCover}}} & \multicolumn{2}{c}{\textbf{\textsc{ArrowCloak}}} & \multicolumn{2}{c}{\textbf{White-box}} & \multirow{2}{*}{\textbf{Black-box}} \\
\cmidrule(lr){3-4}\cmidrule(lr){5-6}\cmidrule(lr){7-8}\cmidrule(lr){9-10}
& & Our Attack & Rel. Black & Our Attack & Rel. Black & Our Attack & Rel. Black & Acc. & Rel. Black & \\
\midrule
\multirow{3}{*}{ViT-Base}
& CIFAR-10   & 98.4\% & $1.55\times$ & 98.4\% & $1.55\times$ & 98.5\% & $1.55\times$ & 98.1\% & $1.55\times$ & 63.4\% \\
& CIFAR-100  & 90.5\% & $1.52\times$ & 90.6\% & $1.52\times$ & 90.0\% & $1.51\times$ & 90.9\% & $1.52\times$ & 59.7\% \\
& FOOD-101   & 87.9\% & $1.49\times$ & 85.6\% & $1.45\times$ & 84.9\% & $1.43\times$ & 83.4\% & $1.41\times$ & 59.2\% \\
\midrule
\multirow{2}{*}{ViT-Large}
& CIFAR-100  & 91.8\% & $1.33\times$ & 91.8\% & $1.33\times$ & 91.1\% & $1.32\times$ & 91.6\% & $1.33\times$ & 68.8\% \\
& FOOD-101   & 88.9\% & $1.36\times$ & 86.9\% & $1.33\times$ & 85.9\% & $1.32\times$ & 85.5\% & $1.31\times$ & 65.3\% \\
\midrule
\multirow{4}{*}{GPT2-Base}
& SST2       & 92.1\% & $1.46\times$ & 92.1\% & $1.46\times$ & 92.6\% & $1.47\times$ & 92.4\% & $1.47\times$ & 63.0\% \\
& QQP        & 88.1\% & $1.34\times$ & 88.1\% & $1.34\times$ & 87.9\% & $1.34\times$ & 88.1\% & $1.34\times$ & 65.6\% \\
& QNLI       & 87.4\% & $1.63\times$ & 87.1\% & $1.62\times$ & 87.0\% & $1.62\times$ & 86.7\% & $1.61\times$ & 53.7\% \\
& MNLI       & 80.8\% & $1.69\times$ & 80.5\% & $1.69\times$ & 80.4\% & $1.69\times$ & 80.8\% & $1.69\times$ & 47.7\% \\
\midrule
\multirow{2}{*}{Llama3.2-1B}
& GSM8K      & 34.0\% & $5.08\times$ & 31.8\% & $4.75\times$ & 30.7\% & $4.58\times$ & 35.5\% & $5.30\times$ & 6.7\% \\
& IFEVAL     & 58.7\% & $4.85\times$ & 55.5\% & $4.59\times$ & 56.7\% & $4.69\times$ & 54.6\% & $4.51\times$ & 12.1\% \\
\midrule
\multirow{2}{*}{Llama3.2-3B}
& GSM8K      & 64.2\% & $2.32\times$ & 61.6\% & $2.22\times$ & 61.9\% & $2.23\times$ & 64.4\% & $2.32\times$ & 27.7\% \\
& IFEVAL     & 74.2\% & $7.07\times$ & 71.7\% & $6.83\times$ & 73.8\% & $7.03\times$ & 73.5\% & $7.00\times$ & 10.5\% \\
\midrule
\multicolumn{2}{l}{\textbf{Average}}
& -- & $\mathbf{2.47\times}$ & -- & $\mathbf{2.44\times}$ & -- & $\mathbf{2.44\times}$ & -- & $2.49\times$ & $1.00\times$ \\
\bottomrule
\end{tabular}}
\end{table}

\subsection{Attack Performance}
\label{sec:attack_performance}

Table~\ref{tab:attack_exp} reports attack accuracy across all evaluated models and datasets. \textsc{TLG} exhibits security vulnerabilities, consistent with prior findings~\cite{wang_game_2025}. More importantly, our attack also reduces \textsc{GroupCover} and \textsc{ArrowCloak}, the stronger unilateral schemes, to near white-box behaviour: their average Rel. Black ratio of $2.44\times$ closely matches the white-box bound at $2.49\times$. The gap to the white-box bound stays small across vision, language understanding, and reasoning tasks, indicating that the leakage we exploit is intrinsic to unilateral obfuscation. 

Figure~\ref{fig:multi_layer_attack} further confirms that attack accuracy against \textsc{GroupCover} and \textsc{ArrowCloak} stays high regardless of the number of protected layers, since each layer is compromised independently and additional protected layers introduce no compounding noise.

\begin{tcolorbox}[
    colback=gray!10,
    colframe=black,
    arc=4pt,
    boxrule=1pt,
    left=8pt,
    right=8pt,
    top=6pt,
    bottom=6pt
]
\textbf{Answer to RQ1:} Our reconstruction attacks substantially weaken both \textsc{GroupCover} and \textsc{ArrowCloak}, increasing the attack accuracy by $2.44\times$ over the black-box baseline.
\end{tcolorbox}

\subsection{Defense Effectiveness}
\label{sec:defense_effectiveness}

\begin{wrapfigure}{r}{0.45\linewidth}
    \centering
    \vspace{-12pt}
    \includegraphics[width=\linewidth]{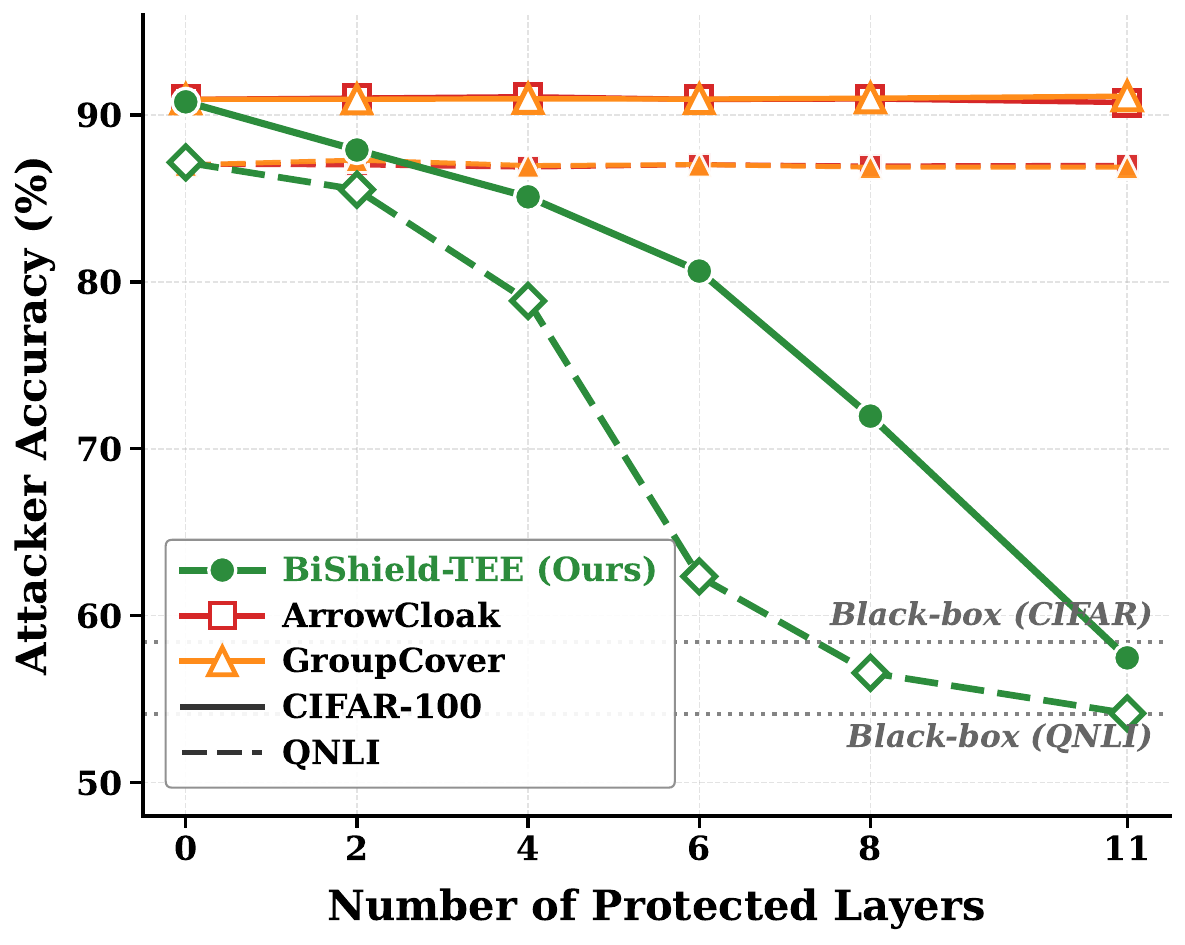}
    \caption{Attack accuracy vs. the number of protected layers on ViT-Base and GPT2-Base.}
    \label{fig:multi_layer_attack}
    \vspace{-12pt}
\end{wrapfigure}


\textbf{Security against adaptive attacks.}
Reconstruction attacks designed for unilateral defenses cannot directly compromise bilateral protection, since recovering only one side leaves the other side missing. We therefore design four attack strategies tailored to the bilateral structure: parameter substitution as the black-box baseline, input-only and output-only recovery, and a joint adaptive attack that alternately fits one side's key with the other fixed, minimizing the residual against $\mathbf{W}_{\text{pre}}$. As shown in Table~\ref{tab:BiShield_defense_analysis}, unilateral recovery causes catastrophic performance collapse: on Llama models, both Input-Only and Output-Only attacks leave the surrogate unable to produce coherent outputs. Even the Joint I\&O attack yields negligible gains, and on most settings the strongest adversary is the parameter substitution baseline itself.

\textbf{Security against the strongest unilateral defense.}
We adopt the most adversary-favorable comparison: for \textsc{BiShield-TEE} we report the highest attack accuracy across all adaptive strategies in Table~\ref{tab:BiShield_defense_analysis}, while for the unilateral baseline we take the lowest attack accuracy in Table~\ref{tab:attack_exp}. Even when comparing our worst case against strongest unilateral defense's best case, Table~\ref{tab:defense_exp} shows that \textsc{BiShield-TEE} reduces Rel. Black to an average of $1.00\times$, against $2.42\times$ for the strongest unilateral defense. 
Its security compounds across layers, as layer-wise reconstruction errors accumulate and progressively diminish attacker capability as shown in Figure~\ref{fig:multi_layer_attack}.

\begin{table}[htbp]
\centering
\scriptsize
\begin{minipage}[t]{0.50\textwidth}
\centering
\setlength{\tabcolsep}{3pt}
\renewcommand{\arraystretch}{1.05}
\caption{Adaptive attack accuracy against \textsc{BiShield-TEE}. Results cover input-only, output-only, joint input\&output recovery and parameter substitution attacks.}
\label{tab:BiShield_defense_analysis}
\begin{tabular}{@{}llcccc@{}}
\toprule
\multirow{2.5}{*}{\textbf{Model}} & \multirow{2.5}{*}{\textbf{Data}} & \multicolumn{4}{c}{\textbf{Attack Strategies on \textsc{BiShield-TEE}}} \\
\cmidrule(l){3-6}
& & \textbf{\makecell{Input\\Only}} & \textbf{\makecell{Output\\Only}} & \textbf{\makecell{Joint\\I\&O}} & \textbf{\makecell{Param.\\Subst.}} \\
\midrule
\multirow{3}{*}{ViT-Base} & C10  & 10.7\% & 14.8\% & 63.6\% & 63.4\% \\
                         & C100 & 1.7\%  & 6.1\%  & 59.7\% & 59.7\% \\
                         & F101 & 1.0\%  & 4.5\%  & 59.0\% & 59.2\% \\
\midrule
\multirow{2}{*}{ViT-Large} & C100 & 2.5\% & 6.7\% & 68.6\% & 68.8\% \\
                          & F101 & 1.5\% & 4.3\% & 65.3\% & 65.3\% \\
\midrule
\multirow{4}{*}{GPT2-Base} & SST2 & 56.2\% & 52.4\% & 63.0\% & 63.0\% \\
                          & QQP  & 63.5\% & 63.4\% & 65.6\% & 65.6\% \\
                          & QNLI & 51.6\% & 50.5\% & 53.7\% & 53.7\% \\
                          & MNLI & 34.3\% & 37.4\% & 47.7\% & 47.7\% \\
\midrule
\multirow{2}{*}{Llama-1B} & GSM8K  & 0.0\% & 0.0\% & 7.0\%  & 6.7\%  \\
                          & IFEVAL & 0.0\% & 0.0\% & 12.0\% & 12.1\% \\
\midrule
\multirow{2}{*}{Llama-3B} & GSM8K  & 0.0\% & 0.0\% & 28.0\% & 27.7\% \\
                          & IFEVAL & 0.0\% & 0.0\% & 9.4\%  & 10.5\% \\
\bottomrule
\end{tabular}
\end{minipage}%
\hfill
\begin{minipage}[t]{0.46\textwidth}
\centering
\setlength{\tabcolsep}{3pt}
\renewcommand{\arraystretch}{1.05}
\caption{Defense effectiveness comparison. ``Prior Best'' is the strongest defense among the unilateral schemes in Table~\ref{tab:attack_exp}; ``Ours'' is the highest attack accuracy achieved by the best adaptive attack strategy in Table~\ref{tab:BiShield_defense_analysis}.}
\label{tab:defense_exp}
\begin{tabular}{@{}llcccc@{}}
\toprule
\multirow{2}{*}{\textbf{Model}} & \multirow{2}{*}{\textbf{Dataset}} & \multicolumn{2}{c}{\textbf{Prior Best }} & \multicolumn{2}{c}{\textbf{Ours}} \\
\cmidrule(lr){3-4}\cmidrule(lr){5-6}
& & Acc. & Rel.Black & Acc. & Rel.Black \\
\midrule
\multirow{3}{*}{ViT-Base}
& C10   & 98.4\% & $1.55\times$ & 63.6\% & $\mathbf{1.00\times}$ \\
& C100  & 90.0\% & $1.51\times$ & 59.7\% & $\mathbf{1.00\times}$ \\
& F101  & 84.9\% & $1.43\times$ & 59.2\% & $\mathbf{1.00\times}$ \\
\midrule
\multirow{2}{*}{ViT-Large}
& C100  & 91.1\% & $1.32\times$ & 68.8\% & $\mathbf{1.00\times}$ \\
& F101  & 85.9\% & $1.32\times$ & 65.3\% & $\mathbf{1.00\times}$ \\
\midrule
\multirow{4}{*}{GPT2-Base}
& SST2  & 92.1\% & $1.46\times$ & 63.0\% & $\mathbf{1.00\times}$ \\
& QQP   & 87.9\% & $1.34\times$ & 65.6\% & $\mathbf{1.00\times}$ \\
& QNLI  & 87.0\% & $1.62\times$ & 53.7\% & $\mathbf{1.00\times}$ \\
& MNLI  & 80.4\% & $1.69\times$ & 47.7\% & $\mathbf{1.00\times}$ \\
\midrule
\multirow{2}{*}{Llama-1B}
& GSM8K & 30.7\% & $4.58\times$ & 7.0\%  & $\mathbf{1.04\times}$ \\
& IFEVAL  & 55.5\% & $4.59\times$ & 12.1\% & $\mathbf{1.00\times}$ \\
\midrule
\multirow{2}{*}{Llama-3B}
& GSM8K & 61.6\% & $2.22\times$ & 28.0\% & $\mathbf{1.01\times}$ \\
& IFEVAL  & 71.7\% & $6.83\times$ & 10.5\% & $\mathbf{1.00\times}$ \\
\midrule
\multicolumn{2}{@{}l}{\textbf{Average}} & -- & $2.42\times$ & -- & $\mathbf{1.00\times}$ \\
\bottomrule
\end{tabular}
\end{minipage}
\vspace{-5pt}
\end{table}

\textbf{Utility preservation.}
Bilateral obfuscation should also preserve service quality for authorized users. Table~\ref{tab:utility} compares the original model with its \textsc{BiShield-TEE} obfuscated version. The matching accuracies across all datasets confirm that the TEE inverse transformations recover the original output and preserve authorized inference quality.

\begin{table}[htbp]
\centering
\scriptsize
\setlength{\tabcolsep}{3pt}
\renewcommand{\arraystretch}{1.05}
\caption{Authorized utility of \textsc{BiShield-TEE}. ``Original'' is the unprotected fine-tuned victim; ``Obfuscated'' is the same model under bilateral obfuscation.}
\label{tab:utility}
\resizebox{\textwidth}{!}{%
\begin{tabular}{@{}lccccccccccccc@{}}
\toprule
\multirow{2}{*}{\textbf{Setting}}
& \multicolumn{3}{c}{\textbf{ViT-Base}}
& \multicolumn{2}{c}{\textbf{ViT-Large}}
& \multicolumn{4}{c}{\textbf{GPT2-Base}}
& \multicolumn{2}{c}{\textbf{Llama3.2-1B}}
& \multicolumn{2}{c}{\textbf{Llama3.2-3B}} \\
\cmidrule(lr){2-4}\cmidrule(lr){5-6}\cmidrule(lr){7-10}\cmidrule(lr){11-12}\cmidrule(lr){13-14}
& C10 & C100 & F101 & C100 & F101 & SST2 & QQP & QNLI & MNLI & GSM8K & IFEVAL & GSM8K & IFEVAL \\
\midrule
Original    & 98.1\% & 90.9\% & 83.4\% & 91.6\% & 85.5\% & 92.4\% & 88.1\% & 86.7\% & 80.8\% & 35.5\% & 54.6\% & 64.4\% & 73.5\% \\
Obfuscated  & 98.1\% & 90.9\% & 83.4\% & 91.6\% & 85.5\% & 92.4\% & 88.1\% & 86.7\% & 80.8\% & 35.5\% & 54.6\% & 64.4\% & 73.5\% \\
\bottomrule
\end{tabular}}
\vspace{-6pt}
\end{table}

\begin{tcolorbox}[
    colback=gray!10,
    colframe=black,
    arc=4pt,
    boxrule=1pt,
    left=8pt,
    right=8pt,
    top=6pt,
    bottom=6pt
]
\textbf{Answer to RQ2:} \textsc{BiShield-TEE} remains secure against all adaptive attacks, consistently reducing adversaries to black-box level performance with Rel. Black averaging $1.00\times$.
\end{tcolorbox}

\subsection{Real-Device Performance}
\label{Real-Device}
We benchmark end-to-end inference latency, with each experiment repeated 10 times after warm-up. As shown in Table~\ref{tab:latency-comparison}, \textsc{BiShield-TEE} achieves the lowest latency on every model, delivering on average $1.70\times$ and $1.46\times$ speedups over \textsc{GroupCover} and \textsc{ArrowCloak}.
This advantage stems from a structural property of bilateral protection. Unilateral schemes apply obfuscation on only one side of the weight matrix and rely on more complex transformations but remain vulnerable to our attacks. In contrast, bilateral protection derives security from joint indistinguishability, allowing each side to remain lightweight while maintaining strong protection. Our results confirm that this design achieves both higher security and lower computational overhead. Further details on the real-device evaluation and recovery-cost analysis are provided in Appendix~\ref{app:experimental-details}.

\begin{table}[htbp]
\caption{End-to-end inference latency (ms) across obfuscation schemes. ``Mean Ratio'' is the average slowdown relative to \textsc{BiShield-TEE} across all models.}

\label{tab:latency-comparison}
\centering
\scriptsize
\setlength{\tabcolsep}{3pt}
\renewcommand{\arraystretch}{1.15}
\resizebox{\linewidth}{!}{%
\begin{tabular}{@{}l cccccc@{}}
\toprule
\textbf{Scheme} & \textbf{ViT-Base} & \textbf{ViT-Large} & \textbf{GPT2-Base} & \textbf{Llama3.2-1B} & \textbf{Llama3.2-3B} & \textbf{\textit{Mean Ratio}} \\
\midrule
\textsc{GroupCover}
& $1667.97 \pm 14.58$
& $4770.45 \pm 30.74$
& $1028.13 \pm 2.29$
& $3995.28 \pm 13.06$
& $9848.23 \pm 74.23$
& $1.70\times$ \\
\textsc{ArrowCloak}
& $1427.14 \pm 2.35$
& $4234.46 \pm 21.99$
& $853.99 \pm 1.47$
& $3386.38 \pm 24.31$
& $8441.53 \pm 32.54$
& $1.46\times$ \\
\midrule
\textbf{\textsc{BiShield-TEE}}
& $\mathbf{1073.74 \pm 11.24}$
& $\mathbf{2918.47 \pm 10.16}$
& $\mathbf{624.04 \pm 11.54}$
& $\mathbf{2272.86 \pm 21.10}$
& $\mathbf{5148.10 \pm 7.22}$
& $\mathbf{1.00\times}$ \\
\bottomrule
\end{tabular}%
}
\vspace{-3pt}
\end{table}

\begin{tcolorbox}[
colback=gray!10,
colframe=black,
arc=4pt,
boxrule=1pt,
left=8pt,
right=8pt,
top=6pt,
bottom=6pt
]
\textbf{Answer to RQ3:} \textsc{BiShield-TEE} achieves lower end-to-end inference latency than the two unilateral schemes across all five models.
\end{tcolorbox}

\section{Discussion}
\label{sec:discussion}


\textbf{Availability of pre-trained models.}
Fine-tuning from public checkpoints has become a prevalent practice, making $\mathbf{W}_{\text{pre}}$ widely available to the adversary.
To assess how strongly our attacks rely on this resource, we vary the magnitude of the fine-tuning increment and measure the resulting surrogate--victim similarity. 
The attacks remain highly effective at similarity levels typical of standard fine-tuning, and their effectiveness only diminishes when the increment becomes sufficiently large to noticeably degrade the victim model itself.
Full results are reported in Appendix~\ref{app:correlation}. 

\paragraph{Ablation of Bilateral Transformation Components.}
We ablate diagonal scaling while retaining the bilateral Householder transformations. The Householder-only variant yields a mean Rel. Black of $1.34\times$, compared with $1.00\times$ for the full design. Householder transformations introduce ambiguity in key recovery, but this alone leaves the attacker with measurable gains over the black-box baseline. Adding diagonal scaling largely removes these gains. The complete results and an analysis of both components are provided in Appendix~\ref{app:transformation-ablation}.

\textbf{GPU TEEs.} Hardware-backed GPU TEEs offer confidential computing in principle, but practical deployment on edge devices remains limited~\cite{wang_confidential_2024}. Existing solutions target server-class platforms such as NVIDIA H100 or require firmware modifications~\cite{volos_graviton_2018,wang_building_2024,StrongBox}, both incompatible with commodity edge GPUs. \textsc{BiShield-TEE} sidesteps this dependence with a purely algorithmic design deployable on off-the-shelf hardware.


\textbf{Limitations.}
Our real-device evaluation is limited to Intel SGX. Other TEEs such as Arm TrustZone~\cite{pinto_demystifying_2019} remain to be tested. The current design does not support quantized inference because INT8/INT4 scaling, clipping and rounding may disrupt the algebraic relations used to recover outputs. Our analysis does not consider side-channel attacks against the TEE~\cite{HyperTheft,CipherH}. Model updates in on-device fine-tuning and federated learning may introduce additional privacy risks that remain for future study.


\section{Conclusions}

We identified SDP as a fundamental vulnerability shared across existing TEE-based unilateral obfuscation defenses. Building on this insight, we developed the first successful attacks against \textsc{GroupCover} and \textsc{ArrowCloak}, boosting attack accuracy by over $2.4\times$ relative to the black-box baseline and revealing an overlooked risk for TEE-based on-device model protection. To address the issue, we proposed \textsc{BiShield-TEE}, a bilateral framework that induces Ambiguous Solutions through joint input-output obfuscation, confining attack accuracy to the black-box baseline while delivering practical efficiency on real hardware. Together, these contributions establish a unified offensive-defensive perspective on TEE-based model protection and provide a deployable foundation for safeguarding proprietary models in real-world edge scenarios.

\begin{ack}
This work is supported by the National Cyber Security-National Science and Technology Major Project (Grant No. 2025ZD1503603).
\end{ack}

\bibliographystyle{unsrt}
\bibliography{reference}

\newpage


\appendix

\section{Details of Existing Obfuscation Schemes}
\label{app:methods}

\textbf{\textsc{GroupCover}.}
\textsc{GroupCover} \cite{zhang_groupcover_2024} partitions the columns of a weight matrix $\mathbf{W}_{\text{vic}} \in \mathbb{R}^{d \times n}$ into disjoint groups of size $s$. Within each group, linear combinations of the $s$ columns are applied, yielding a dense $s \times s$ transformation submatrix. Since interactions are restricted within groups, the overall transformation matrix exhibits a block-diagonal structure. To conceal this structure, random row and column permutations are applied. The resulting transformation can be expressed as
\begin{equation}
\mathbf{T}_{\textsc{GroupCover}} = \mathbf{P}_1
\begin{bmatrix}
\mathbf{T}_1 & \mathbf{0} & \cdots & \mathbf{0} \\
\mathbf{0} & \mathbf{T}_2 & \cdots & \mathbf{0} \\
\vdots & \vdots & \ddots & \vdots \\
\mathbf{0} & \mathbf{0} & \cdots & \mathbf{T}_k
\end{bmatrix}
\mathbf{P}_2,
\quad
\mathbf{T}_i \in \mathbb{R}^{s \times s},
\end{equation}
where $\mathbf{P}_1, \mathbf{P}_2 \in \mathbb{R}^{n \times n}$ are permutation matrices and each $\mathbf{T}_i$ corresponds to a feature group.

\vspace{0.5em}
\textbf{\textsc{ArrowCloak}.}
\textsc{ArrowCloak} \cite{wang_game_2025} applies column-wise scaling and a shared additive masking to $\mathbf{W}_{\text{vic}} \in \mathbb{R}^{d \times n}$. The obfuscated matrix is given by
\begin{equation}
\mathbf{W}_{\text{obf}} = (\mathbf{W}_{\text{vic}} \mathbf{D}_p + \mathbf{v}\mathbf{q}^{\top}) \mathbf{\Pi},
\end{equation}
where $\mathbf{D}_p = \operatorname{diag}(p_1, \dots, p_n)$ defines column-wise scaling, $\mathbf{q} \in \mathbb{R}^n$ controls the additive coefficients, and $\mathbf{\Pi}$ is a permutation matrix. The shared mask vector $\mathbf{v}$ is constructed as a linear combination of columns:
\begin{equation}
\mathbf{v} = \mathbf{W}_{\text{vic}} \mathbf{k},
\end{equation}
where $\mathbf{k} \in \mathbb{R}^n$ is a secret coefficient vector. Substituting $\mathbf{v}$ yields the equivalent transformation
\begin{equation}
\mathbf{T}_{\textsc{ArrowCloak}} = \bigl(\operatorname{diag}(p_1, \dots, p_n) + \mathbf{k}\mathbf{q}^{\top}\bigr)\mathbf{\Pi},
\end{equation}
which can be explicitly written as
\begin{equation}
\label{eq:arrowcloak-T}
\mathbf{T}_{\textsc{ArrowCloak}} =
\begin{bmatrix}
p_1 + k_1 q_1 & k_1 q_2 & \cdots & k_1 q_n \\
k_2 q_1 & p_2 + k_2 q_2 & \cdots & k_2 q_n \\
\vdots & \vdots & \ddots & \vdots \\
k_n q_1 & k_n q_2 & \cdots & p_n + k_n q_n
\end{bmatrix}\mathbf{\Pi}.
\end{equation}

\section{Proof of Parameter Recovery for \textsc{GroupCover}}
\label{groupcover_proof}

We show that \textsc{GroupCover}'s reliance on an $s \times s$ obfuscation matrix to protect a $d \times s$ target matrix is inherently vulnerable when $s \ll d$: once the grouping is compromised, the recovered surrogate $\mathbf{W}_{\text{sur}}$ can be made close to the victim $\mathbf{W}_{\text{vic}}$.

\begin{proof}
Let $\mathbf{W}_{\text{pre}}, \mathbf{W}_{\text{vic}}, \mathbf{W}_{\text{obf}} \in \mathbb{R}^{d \times s}$, and write
\[
\mathbf{W}_{\text{vic}} = \mathbf{W}_{\text{pre}} + \Delta, \qquad \mathbf{W}_{\text{obf}} = \mathbf{W}_{\text{vic}} \mathbf{T},
\]
where $\mathbf{T} \in \mathbb{R}^{s \times s}$ is the unknown invertible obfuscation matrix. We assume $\mathbf{W}_{\text{pre}}$ has full column rank (which is generic for $s \ll d$) and that $\Delta$ is a small perturbation, i.e.\ $\|\Delta\| \ll \|\mathbf{W}_{\text{pre}}\|$.

\paragraph{Estimating $\mathbf{T}$.}
The attacker solves
\[
\hat{\mathbf{T}} = \arg\min_{\mathbf{X} \in \mathbb{R}^{s \times s}} \| \mathbf{W}_{\text{pre}} \mathbf{X} - \mathbf{W}_{\text{obf}} \|_F^2,
\]
which admits the unique solution $\hat{\mathbf{T}} = \mathbf{W}_{\text{pre}}^{\dagger}\mathbf{W}_{\text{obf}}$ with $\mathbf{W}_{\text{pre}}^{\dagger} = (\mathbf{W}_{\text{pre}}^\top \mathbf{W}_{\text{pre}})^{-1} \mathbf{W}_{\text{pre}}^\top$. Substituting $\mathbf{W}_{\text{obf}} = (\mathbf{W}_{\text{pre}} + \Delta)\mathbf{T}$ and using $\mathbf{W}_{\text{pre}}^{\dagger}\mathbf{W}_{\text{pre}} = \mathbf{I}_s$,
\[
\hat{\mathbf{T}} = (\mathbf{I}_s + \mathbf{M})\mathbf{T}, \qquad \mathbf{M} := \mathbf{W}_{\text{pre}}^{\dagger}\Delta.
\]
Since $\|\mathbf{M}\| \ll 1$, $\mathbf{I}_s + \mathbf{M}$ is invertible and $\hat{\mathbf{T}}^{-1} = \mathbf{T}^{-1}(\mathbf{I}_s + \mathbf{M})^{-1}$.

\paragraph{First-order expansion of $\mathbf{W}_{\text{sur}}$.}
The surrogate $\mathbf{W}_{\text{sur}} := \mathbf{W}_{\text{obf}}\hat{\mathbf{T}}^{-1}$ satisfies
\[
\mathbf{W}_{\text{sur}} = (\mathbf{W}_{\text{pre}} + \Delta)(\mathbf{I}_s + \mathbf{M})^{-1}.
\]
Applying the Neumann series $(\mathbf{I}_s + \mathbf{M})^{-1} = \mathbf{I}_s - \mathbf{M} + O(\|\mathbf{M}\|^2)$ and discarding terms of order $\|\Delta\|^2$ or higher,
\[
\mathbf{W}_{\text{sur}} - \mathbf{W}_{\text{vic}} = -\mathbf{P}\Delta + O(\|\Delta\|^2),
\]
where $\mathbf{P} := \mathbf{W}_{\text{pre}}\mathbf{W}_{\text{pre}}^{\dagger}$ is the orthogonal projector onto $\operatorname{col}(\mathbf{W}_{\text{pre}})$, an $s$-dimensional subspace of $\mathbb{R}^d$.

\paragraph{Error scales as $s/d$.}
The reconstruction error is thus the projection of $\Delta$ onto an $s$-dimensional subspace of $\mathbb{R}^d$. Since the column space of $\mathbf{W}_{\text{pre}}$ occupies only a small fraction of the ambient space when $s \ll d$, a generic perturbation $\Delta$ has roughly an $s/d$ fraction of its energy lying in this subspace, giving
\[
\|\mathbf{W}_{\text{sur}} - \mathbf{W}_{\text{vic}}\|_F^2 \;\approx\; \frac{s}{d}\,\|\Delta\|_F^2.
\]
When $s \ll d$, the reconstruction error is therefore much smaller than the original perturbation, so $\mathbf{W}_{\text{sur}}$ closely recovers $\mathbf{W}_{\text{vic}}$.
\end{proof}

\section{Detailed Attack Analysis}
\label{app:attack}

This appendix provides additional diagnostics and complexity analysis for the attacks on \textsc{GroupCover} and \textsc{ArrowCloak}.

\subsection{Details of the \textsc{GroupCover} Attack}
\paragraph{Preprocessing of the estimated transformation.}
The attack starts from the least-squares estimate
\begin{equation}
\hat{\mathbf{T}} = \mathbf{W}_{\text{pre}}^{\dagger} \mathbf{W}_{\text{obf}}.
\end{equation}
Although $\hat{\mathbf{T}}$ already preserves the block-sparse structure of the true transformation, its raw entries may vary substantially across rows and columns due to scale differences. To sharpen the hidden grouping pattern, we normalize each entry by the geometric mean of its row and column energy:
\begin{equation}
\tilde{T}_{ij} = \frac{\hat{T}_{ij}^2}{\sqrt{\sum_k \hat{T}_{ik}^2 \cdot \sum_k \hat{T}_{kj}^2}}.
\label{eq:gc-normalize}
\end{equation}
This preprocessing suppresses weak noisy entries while amplifying the dominant intra-group correlations. After normalization, each row and column exhibits approximately $s$ large entries, which makes the hidden group structure much easier to identify.

\paragraph{Group recovery via greedy clique extraction.}
After normalization, columns belonging to the same group share a common support of $s$ dominant indices, while columns from different groups have nearly disjoint supports. We exploit this by constructing a soft co-support similarity matrix $\mathbf{W} = \tilde{\mathbf{S}}^{\top}\tilde{\mathbf{S}}$ (with the diagonal zeroed out), where $\tilde{\mathbf{S}}$ retains the top-$M$ entries per column of $\tilde{\mathbf{T}}$, with $M = \min(n,\, 4s)$. Each entry $W_{jk}$ thus measures how strongly columns $j$ and $k$ co-activate on the same rows, and groups correspond to dense $s$-cliques in $\mathbf{W}$. We recover them in batches: each unassigned column proposes a candidate clique with its top $s{-}1$ neighbors, scored by the total intra-clique affinity, and candidates are committed in decreasing order of score while skipping any whose columns have already been claimed. The corresponding column group is taken directly as the clique itself. The full procedure is summarized in Algorithm~\ref{alg:gc-group-recovery}.

\begin{algorithm}[t]
\caption{Group Recovery for \textsc{GroupCover}}
\label{alg:gc-group-recovery}
\begin{algorithmic}[1]
\Require Estimated transformation $\hat{\mathbf{T}} \in \mathbb{R}^{n \times n}$, group size $s$
\Ensure Column groups $\mathcal{G}$ and column-to-group assignment $\pi$
\State $\tilde{\mathbf{T}} \gets \textsc{Normalize}(\hat{\mathbf{T}})$ via Eq.~\eqref{eq:gc-normalize};\ \ $\tilde{\mathbf{S}} \gets \textsc{TopM}(\tilde{\mathbf{T}}),\ M = \min(n,\, 4s)$ \Comment{top-$M$ entries per column}
\State $\mathbf{W} \gets \tilde{\mathbf{S}}^{\top}\tilde{\mathbf{S}}$ with $W_{jj}{\gets}0$;\ \ $U \gets [n]$;\ \ $\mathcal{G} \gets \emptyset$
\While{$|U| \geq s$}
    \For{$j \in U$}
        \State $\mathcal{C}_j \gets \{j\} \cup \textsc{TopNeighbors}_{U}(\mathbf{W}[:,j],\, s{-}1)$;\ \ $\mathrm{score}(j) \gets \sum_{u,v\in \mathcal{C}_j} W_{uv}$
    \EndFor
    \For{$j \in U$ in decreasing order of $\mathrm{score}(j)$}
        \If{$\mathcal{C}_j \subseteq U$}
            \State $\pi(c) \gets \mathcal{C}_j\ \ \forall c \in \mathcal{C}_j$;\ \ $\mathcal{G} \gets \mathcal{G} \cup \{\mathcal{C}_j\}$
            \State $U \gets U \setminus \mathcal{C}_j$
        \EndIf
    \EndFor
\EndWhile
\State \Return $\mathcal{G}, \pi$
\end{algorithmic}
\end{algorithm}

\begin{figure}[htbp]
  \centering
  \begin{subfigure}[t]{0.48\linewidth}
    \centering
    \includegraphics[width=\linewidth]{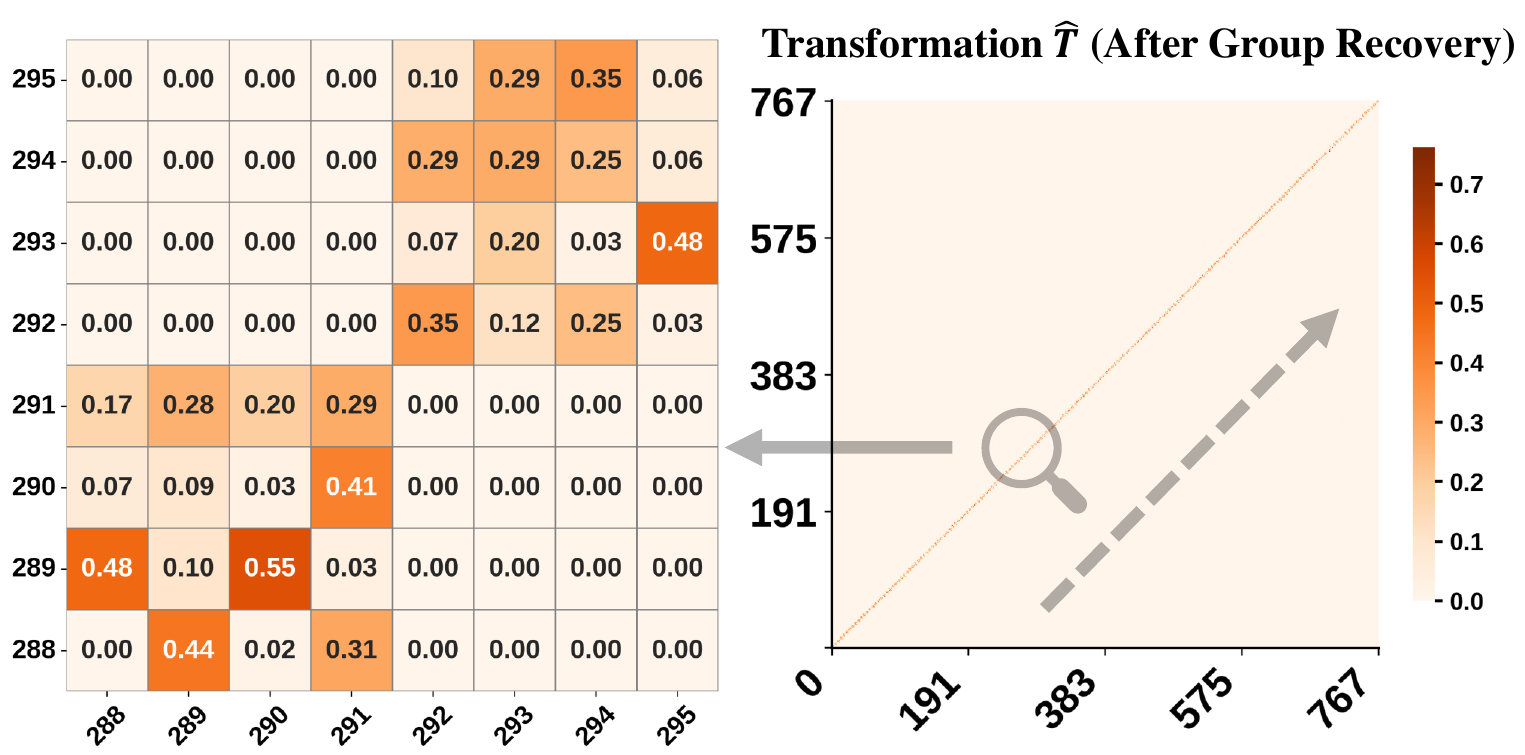}
    \caption{\textsc{GroupCover}: block sparsity in $\hat{\mathbf{T}}$ after preprocessing.}
    \label{fig:heatmap}
  \end{subfigure}\hfill
  \begin{subfigure}[t]{0.48\linewidth}
    \centering
    \includegraphics[width=\linewidth]{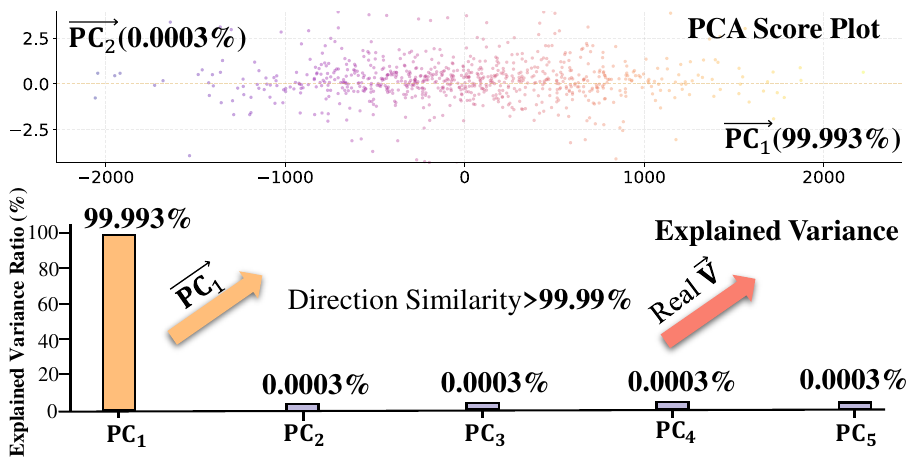}
    \caption{\textsc{ArrowCloak}: dominant principal component.}
    \label{fig:pca}
  \end{subfigure}
  \caption{Diagnostic visualization of unilateral obfuscation leakage.}
  \label{fig:attack_diagnostics}
\end{figure}

\paragraph{Visualization.}
Figure~\ref{fig:heatmap} shows the preprocessed transformation matrix for the first-layer $\mathbf{W}_Q$ parameters of ViT-Base. The dominant entries are sharply concentrated in block-diagonal regions, directly revealing the grouping structure used by \textsc{GroupCover}. This confirms that the structural sparsity of the true secret key is preserved in the estimated transformation.

\paragraph{Complexity.}
For a target matrix of size $d \times n$, computing $\hat{\mathbf{T}}$ costs $O(nd \cdot \min(n,d))$. The preprocessing in Eq.~\eqref{eq:gc-normalize} is linear in the size of $\hat{\mathbf{T}}$ and thus costs $O(n^2)$. Constructing $\mathbf{W}$ from the sparsified $\tilde{\mathbf{S}}$ costs $O(nM^2)$, and each iteration of the clique extraction loop touches at most $O(ns)$ entries; with $n/s$ iterations the total clique-recovery cost is $O(n^2)$. The subsequent reconstruction stage has complexity $O(n^2 d)$ when the group size $s$ is a small fixed constant. Therefore, the overall complexity is dominated by $O(n^2 d)$.

\subsection{Details of the \textsc{ArrowCloak} Attack}

\paragraph{Attack overview.}
\textsc{ArrowCloak} obfuscates each weight matrix by combining per-column scaling, a shared additive mask $\mathbf{v}$, and a column permutation $\boldsymbol{\Pi}$ (Eq.~\eqref{eq:arrowcloak-T}). The shared mask is the central source of leakage: because it is added to \emph{every} column, it dominates the column-wise variance of $\mathbf{W}_{\text{obf}}$ and emerges as the leading principal component after centering. We exploit this leakage in three stages, summarized in Algorithm~\ref{alg:ac-recover}: (i) extract the mask direction $\mathbf{v}_{\text{est}}$ via SVD on the centered matrix; (ii) project $\mathbf{W}_{\text{obf}}$ onto the orthogonal complement of $\mathbf{v}_{\text{est}}$ to remove the additive component, then recover the column permutation by Hungarian matching against $\mathbf{W}_{\text{pre}}$; (iii) for each column, jointly fit the scaling coefficient and residual mask weight via least squares.

\begin{algorithm}[t]
\caption{Reconstruction Attack on \textsc{ArrowCloak}}
\label{alg:ac-recover}
\begin{algorithmic}[1]
\Require Obfuscated weights $\mathbf{W}_{\text{obf}} \in \mathbb{R}^{d \times n}$, pre-trained reference $\mathbf{W}_{\text{pre}} \in \mathbb{R}^{d \times n}$
\Ensure Surrogate weights $\mathbf{W}_{\text{sur}}$
\State \textbf{// Stage 1: Mask extraction via PCA}
\State $\mathbf{W}_c \gets \mathbf{W}_{\text{obf}} - \bar{\mathbf{w}}\mathbf{1}_n^{\top}$,\ \ where $\bar{\mathbf{w}}$ is the column-mean vector of $\mathbf{W}_{\text{obf}}$
\State $\mathbf{v}_{\text{est}} \gets$ leading left singular vector of $\mathbf{W}_c$
\State \textbf{// Stage 2: Denoised column alignment}
\State $\mathbf{W}_{\text{clean}} \gets \mathbf{W}_{\text{obf}} - \mathbf{v}_{\text{est}}(\mathbf{v}_{\text{est}}^{\top}\mathbf{W}_{\text{obf}})$
\State $\mathbf{C} \gets \mathbf{1} - \widehat{\mathbf{W}}_{\text{pre}}^{\top}\widehat{\mathbf{W}}_{\text{clean}}$,\ \ where $\widehat{(\cdot)}$ denotes column-wise $\ell_2$ normalization
\State $\boldsymbol{\pi} \gets \textsc{HungarianMatch}(\mathbf{C})$;\ \ $\widetilde{\mathbf{W}}_{\text{obf}} \gets \mathbf{W}_{\text{obf}}[:, \boldsymbol{\pi}]$
\State \textbf{// Stage 3: Per-column least-squares reconstruction}
\For{$i = 1, \ldots, n$}
    \State $\mathbf{B}_i \gets [\widetilde{\mathbf{w}}_{\text{obf}}^{(i)}\ \,\mathbf{v}_{\text{est}}] \in \mathbb{R}^{d \times 2}$
    \State $(\hat{\alpha}_i, \hat{\beta}_i)^{\top} \gets \arg\min_{\boldsymbol{\theta}} \|\mathbf{B}_i \boldsymbol{\theta} - \mathbf{w}_{\text{pre}}^{(i)}\|_2^2$
    \State $\mathbf{w}_{\text{sur}}^{(i)} \gets \hat{\alpha}_i \widetilde{\mathbf{w}}_{\text{obf}}^{(i)} + \hat{\beta}_i \mathbf{v}_{\text{est}}$
\EndFor
\State \Return $\mathbf{W}_{\text{sur}}$
\end{algorithmic}
\end{algorithm}

\paragraph{Visualization.}
As shown in Figure~\ref{fig:pca}, the obfuscated matrix exhibits a clear spectral outlier: the first principal component dominates the spectrum and aligns with the shared mask direction, with a cosine similarity to the true mask exceeding $99.99\%$ across all evaluated layers.

\paragraph{Complexity.}
For a target matrix of size $d \times n$, PCA on the centered matrix costs $O(nd \cdot \min(n,d))$, alignment via the Hungarian algorithm costs $O(n^3)$ in the worst case, and the per-column least-squares stage solves $n$ independent $d \times 2$ systems in $O(nd)$ total. In practice, the distinct directionality of column vectors facilitates efficient matching, and runtime is dominated by the PCA step, making the overall complexity lower than that of attacking \textsc{GroupCover}.

\section{Security Analysis: Ambiguous Solutions}
\label{app:security}

Under bilateral obfuscation,
\begin{equation}
\mathbf{W}_{\mathrm{obf}} = \mathbf{T}_l\,\mathbf{W}_{\mathrm{vic}}\,\mathbf{T}_r,
\end{equation}
where $\mathbf{T}_l = H(\mathbf{u}_\star)\mathbf{D}_l^\star$ and $\mathbf{T}_r = \mathbf{D}_r^\star G(\mathbf{v}_\star)$ are the secret transformations. Here, $H(\mathbf{u})=\mathbf{I}_d-2\mathbf{u}\mathbf{u}^\top$ and $G(\mathbf{v})=\mathbf{I}_n-2\mathbf{v}\mathbf{v}^\top$ are Householder reflections with $\mathbf{u}\in S^{d-1}$ and $\mathbf{v}\in S^{n-1}$, and the invertible diagonal factors satisfy
\[
\kappa^{-1}\le |(\mathbf{D}_l)_{ii}|\le\kappa, \qquad
\kappa^{-1}\le |(\mathbf{D}_r)_{jj}|\le\kappa,
\]
for $\kappa>1$. Let $\mathcal{P}$ denote the resulting key space and write
\[
F(\mathbf{u},\mathbf{D}_l,\mathbf{D}_r,\mathbf{v}) = \mathbf{D}_l^{-1}H(\mathbf{u})\,\mathbf{W}_{\mathrm{obf}}\,G(\mathbf{v})\mathbf{D}_r^{-1}
\]
for the recovery map. At the true key $p_\star=(\mathbf{u}_\star,\mathbf{D}_l^\star,\mathbf{D}_r^\star,\mathbf{v}_\star)$, this map returns $\mathbf{W}_{\mathrm{vic}}$. We consider a fine-tuning increment $\boldsymbol{\Delta}=\mathbf{W}_{\mathrm{vic}}-\mathbf{W}_{\mathrm{pre}}$ satisfying $\|\boldsymbol{\Delta}\|_F\le\varepsilon\ll\|\mathbf{W}_{\mathrm{vic}}\|_F$. The adversary observes $(\mathbf{W}_{\mathrm{obf}},\mathbf{W}_{\mathrm{pre}})$ and seeks a surrogate $\widehat{\mathbf{W}}$ that retains the victim's fine-tuned functionality.

\subsection{The Adversary's Recovery Problem}

Given $(\mathbf{W}_{\mathrm{obf}},\mathbf{W}_{\mathrm{pre}})$, a natural residual-based recovery strategy is to search for a key $p\in\mathcal{P}$ that makes $F(p)$ close to $\mathbf{W}_{\mathrm{pre}}$:
\begin{equation}
p^\dagger \in \arg\min_{p\in\mathcal{P}} \|F(p)-\mathbf{W}_{\mathrm{pre}}\|_F.
\label{eq:adv-obj}
\end{equation}
Define the admissible set at residual tolerance $\eta$,
\[
U_\eta = \bigl\{p\in\mathcal{P} : \|F(p)-\mathbf{W}_{\mathrm{pre}}\|_F\le\eta\bigr\},
\]
which contains all keys whose reconstructed weights satisfy the pre-trained reference constraint at tolerance $\eta$. The true key satisfies
\[
r_\star := \|F(p_\star)-\mathbf{W}_{\mathrm{pre}}\|_F = \|\boldsymbol{\Delta}\|_F \le \varepsilon,
\]
so $p_\star\in U_\varepsilon$. Every key in $U_\varepsilon$ satisfies the same residual bound as the true key, although its residual need not be smaller than $r_\star$.

\subsection{Ambiguity of the Transformation Keys}

The recovery map $F$ admits exact symmetries that produce distinct keys with identical recovered weights. The simplest is the antipodal symmetry of Householder reflections:
\[
H(-\mathbf{u}) = H(\mathbf{u}), \qquad G(-\mathbf{v}) = G(\mathbf{v}).
\]
Hence the four keys
\[
(\pm\mathbf{u}_\star,\,\mathbf{D}_l^\star,\,\mathbf{D}_r^\star,\,\pm\mathbf{v}_\star)
\]
all yield identical $F$-values and identical residuals.

Bilateral obfuscation also admits reciprocal scaling between the two diagonal factors. For any $c>0$ such that the rescaled factors remain admissible, define
\[
p_c = \left(\mathbf{u}_\star,\,c\mathbf{D}_l^\star,\,c^{-1}\mathbf{D}_r^\star,\,\mathbf{v}_\star\right).
\]
Then
\[
F(p_c) = (c\mathbf{D}_l^\star)^{-1}H(\mathbf{u}_\star)\mathbf{W}_{\mathrm{obf}}G(\mathbf{v}_\star)(c^{-1}\mathbf{D}_r^\star)^{-1} = F(p_\star).
\]
When the magnitudes of the true diagonal entries lie strictly between $\kappa^{-1}$ and $\kappa$, all $c$ sufficiently close to $1$ are valid. Thus, a fixed recovered weight matrix can correspond to a continuum of distinct transformation pairs. These symmetries establish non-uniqueness of the transformation keys, although they do not by themselves produce different surrogate weights.

\subsection{A Continuum of Admissible Reconstructions}

We next show that approximate agreement with $\mathbf{W}_{\mathrm{pre}}$ can admit distinct recovered weight matrices. Choose a nonzero row of $\mathbf{W}_{\mathrm{vic}}$, indexed by $i$, and let $\mathbf{e}_i$ denote the corresponding standard basis vector. Define the diagonal perturbation
\[
\mathbf{A}(t) = \mathbf{I}_d + (e^t-1)\mathbf{e}_i\mathbf{e}_i^\top
\]
and the candidate key
\[
p(t) = \left(\mathbf{u}_\star,\,\mathbf{D}_l^\star\mathbf{A}(t),\,\mathbf{D}_r^\star,\,\mathbf{v}_\star\right).
\]
Because $\kappa>1$, $t$ can vary over a nontrivial admissible interval containing $0$, possibly on only one side of $0$ when the corresponding diagonal magnitude lies at a boundary. Substituting into the recovery map gives
\[
F(p(t)) = \mathbf{A}(t)^{-1}\mathbf{W}_{\mathrm{vic}},
\]
and consequently
\[
\|F(p(t))-\mathbf{W}_{\mathrm{vic}}\|_F = |e^{-t}-1|\,\|\mathbf{e}_i^\top\mathbf{W}_{\mathrm{vic}}\|_2.
\]
For $t\ne 0$, the recovered weights differ from $\mathbf{W}_{\mathrm{vic}}$. Nevertheless, each candidate remains exactly consistent with the observed obfuscated matrix under its corresponding transformations.

By the triangle inequality,
\[
\|F(p(t))-\mathbf{W}_{\mathrm{pre}}\|_F \le r_\star + |e^{-t}-1|\,\|\mathbf{e}_i^\top\mathbf{W}_{\mathrm{vic}}\|_2.
\]
Since the second term tends to zero as $t\to 0$, every tolerance $\eta>r_\star$ admits a continuum of such candidates in $U_\eta$. In particular, for every $\delta>0$, sufficiently small admissible perturbations belong to $U_{\varepsilon+\delta}$. If the prior bound has positive slack, $r_\star<\varepsilon$, a continuum also lies within $U_\varepsilon$ itself.

Thus, the same obfuscated matrix can be compatible with distinct reconstructed weights that all satisfy an approximate pre-trained reference constraint. Their residuals need not be equal, but membership in the admissible set alone does not identify the victim weights.

\subsection{Implications for Reconstruction Attacks}

For \textsc{BiShield-TEE}, reconstruction requires jointly estimating the input- and output-side transformations, whose effects are coupled and can compensate for one another. The constructions above establish non-unique key representations and, when the residual tolerance exceeds $r_\star$, multiple admissible weight reconstructions. Moreover, the objective in \eqref{eq:adv-obj} measures agreement with $\mathbf{W}_{\mathrm{pre}}$, so a smaller residual does not necessarily imply better recovery of the private fine-tuning increment. If $\mathbf{W}_{\mathrm{pre}}\in F(\mathcal{P})$, a zero-residual solution reproduces the public checkpoint itself. These observations explain why pre-trained reference fitting alone cannot reliably identify the genuine reconstruction and why useful recovery requires resolving the coupled transformations using additional information.

\section{Experimental Details}
\label{appendix:setup}

\subsection{Datasets and Settings}
Table~\ref{tab:dataset_sizes} summarizes the size of the original training set used to fine-tune the victim model and the size of the surrogate retraining set. Surrogate labels are obtained by querying the victim, and the surrogate set never exceeds $\sim$$1\%$ of the corresponding victim training set.
For the Llama3.2 series, we apply obfuscation only to the intermediate transformer layers, leaving components such as the embedding and output projection untouched. Under this design, the reconstructed surrogate already achieves strong attack performance without retraining, and we therefore evaluate it directly using the lm-evaluation-harness~\cite{eval-harness}. On GSM8K, we use 8-shot prompting and report the \emph{strict-match} accuracy. On IFEval, we use 0-shot prompting and report the average of the \emph{Prompt-Strict} and \emph{Instruction-Strict} metrics.

\begin{table}[htbp]
\centering
\caption{Dataset statistics. \emph{Victim Train} is the size of the dataset used for victim fine-tuning; \emph{Surrogate Train} is the size of the query-answer set used to retrain the surrogate's classification head.}
\label{tab:dataset_sizes}
\small
\begin{tabular}{llrrr}
\toprule
\textbf{Modality} & \textbf{Dataset} & \textbf{Victim Train} & \textbf{Surrogate Train} & \textbf{Ratio} \\
\midrule
\multirow{3}{*}{Vision} & CIFAR-10  & 50{,}000  & 20   & 0.04\% \\
                        & CIFAR-100 & 50{,}000  & 200  & 0.40\% \\
                        & Food-101  & 75{,}750  & 379  & 0.50\% \\
\midrule
\multirow{4}{*}{Language} & SST-2 & 67{,}349  & 134  & 0.20\% \\
                          & QQP   & 363{,}846 & 181  & 0.05\% \\
                          & QNLI  & 104{,}743 & 273  & 0.26\% \\
                          & MNLI  & 392{,}702 & 982  & 0.25\% \\
\bottomrule
\end{tabular}
\end{table}

\subsection{Experimental Pipeline}

\paragraph{Victim Model Training.}
All victim models are obtained by full-parameter fine-tuning from public pre-trained checkpoints using AdamW, with detailed hyperparameters summarized in Table~\ref{tab:hyperparams}. The checkpoint achieving the highest validation accuracy is selected as the victim model. Gradient accumulation steps are set to $1$ throughout all experiments.
For the Llama3.2 series, we adopt a more realistic threat model in which the publicly available base model serves as the attacker's starting point, while the instruction-tuned model serves as the victim. This reflects real-world scenarios where providers apply proprietary alignment pipelines (e.g., SFT and RLHF) on private data to produce capable instruction-following models~\cite{grattafiori_llama_2024}, and naturally captures the capability gap between base and instruction-tuned variants.

\paragraph{Obfuscation and Attack.}
Victim model parameters are obfuscated under each defense scheme and exposed to the adversary. Obfuscation targets the transformer layers, which constitute the majority of model capacity. For unilateral baselines, the adversary applies the attacks described in Section~\ref{sec:attacks}: sparsity-driven reconstruction for \textsc{GroupCover} and PCA-based reconstruction for \textsc{ArrowCloak}.

We analyze a range of attack strategies against \textsc{BiShield-TEE}. \emph{Unilateral recovery} targets a single side of the key, while \emph{adaptive joint recovery} alternately refines the input and output side keys to minimize the gap between the reconstructed weights $W_{\text{sur}}$ and the pre-trained weights $W_{\text{pre}}$, with the number of iterations fixed to $5$; additional iterations only drive $W_{\text{sur}}$ toward $W_{\text{pre}}$ rather than the true $W_{\text{vic}}$. Learning-based recovery offers no advantage in this setting: it lacks an informative initialization and needs to contend with indistinguishable keys (Appendix~\ref{app:security}), so gradient-based optimization may converge to a solution that remains far from the true key. Query-based attacks are likewise ineffective, since the low-rank OTP refreshes masks at each forward pass, leaving intermediate states as inconsistent masked views from which no consistent information about the underlying model can be accumulated.


\paragraph{Surrogate Model Training.}
Since the pre-trained classification head cannot directly adapt to downstream tasks, we retrain only the classification head of the surrogate model. This component contains few parameters and can be effectively trained using the small query-answer sets reported in Table~\ref{tab:dataset_sizes}. To simulate realistic attack conditions, retraining data is drawn from a held-out split disjoint from the victim's validation set, with labels obtained by querying the victim. Surrogate hyperparameters are reported in the lower block of Table~\ref{tab:hyperparams}.
For the Llama3.2 series, the reconstructed surrogate already achieves strong attack performance without further adaptation; we therefore report its accuracy directly after parameter reconstruction.

\begin{table}[htbp]
\centering
\caption{Training hyperparameters for victim fine-tuning and surrogate classification-head retraining. All experiments use the AdamW optimizer with gradient accumulation steps set to $1$. For the surrogate, the learning rate is selected via grid search.}
\label{tab:hyperparams}
\small
\begin{tabular}{llccc}
\toprule
\textbf{Stage} & \textbf{Model} & \textbf{Learning Rate} & \textbf{Batch Size} & \textbf{Epochs} \\
\midrule
\multirow{3}{*}{Victim}    & ViT-Base   & $1\times10^{-4}$ & 128 & 5 \\
                           & ViT-Large  & $1\times10^{-4}$ & 128 & 3 \\
                           & GPT2-Base  & $2\times10^{-5}$ & 128 & 3 \\
\midrule
\multirow{2}{*}{Surrogate} & ViT        & $\{2,5,8\}\times10^{-4},\, 1\times10^{-3}$ & 32 & 10 \\
                           & GPT2-Base  & $\{2,5,8\}\times10^{-4},\, 1\times10^{-3}$ & 16 & 10 \\
\bottomrule
\end{tabular}
\end{table}



\subsection{Real-Device Latency Benchmark.}
\label{app:experimental-details}
For the end-to-end inference experiments in Section~\ref{Real-Device}, we use the fixed input configurations in Table~\ref{tab:latency_config}. Here $B$ denotes the batch size and $l$ the sequence length. For ViT, $l=197$ consists of $196$ image patches and the \texttt{[CLS]} token under a $224\times224$ input with $16\times16$ patches. Each reported latency is the mean and standard deviation of 10 complete forward passes after warm-up.

\begin{table}[htbp]
\centering
\caption{Input configurations for end-to-end inference latency measurement.}
\label{tab:latency_config}
\small
\begin{tabular}{lcc}
\toprule
\textbf{Model} & \textbf{Batch Size $B$} & \textbf{Sequence Length $l$} \\
\midrule
ViT-Base      & 1 & 197 \\
ViT-Large     & 1 & 197 \\
GPT2-Base     & 2 & 64  \\
Llama3.2-1B   & 2 & 64  \\
Llama3.2-3B   & 2 & 64  \\
\bottomrule
\end{tabular}
\end{table}

\paragraph{TEE-Resident Key Footprint.}
Table~\ref{tab:key_footprint} reports the transformation keys retained inside the TEE for output recovery. This is the scheme-specific key footprint rather than the total memory usage of the enclave. \textsc{BiShield-TEE} has the smallest footprint on every model. Averaged across the five models, its footprint is $19.0\%$ smaller than \textsc{ArrowCloak} and $45.5\%$ smaller than \textsc{GroupCover}.

\begin{table}[htbp]
\centering
\caption{TEE-resident transformation-key footprint (MiB). Lower is better.}
\label{tab:key_footprint}
\small
\setlength{\tabcolsep}{5pt}
\begin{tabular}{@{}lccccc@{}}
\toprule
\textbf{Scheme} & \textbf{ViT-Base} & \textbf{ViT-Large}
& \textbf{GPT2-Base} & \textbf{Llama3.2-1B} & \textbf{Llama3.2-3B} \\
\midrule
\textsc{GroupCover}  & 3.80 & 10.13 & 3.80 & 11.25 & 24.94 \\
\textsc{ArrowCloak}  & 2.53 & 6.75  & 2.53 & 7.69  & 17.06 \\
\textbf{\textsc{BiShield-TEE}}
                     & \textbf{2.06} & \textbf{5.39} & \textbf{2.00}
                     & \textbf{6.33} & \textbf{14.03} \\
\bottomrule
\end{tabular}
\end{table}

\paragraph{Recovery-Cost Analysis.}
Let $X\in\mathbb{R}^{l\times d}$ and $Y',Y\in\mathbb{R}^{l\times n}$ denote the input, outsourced output, and recovered output of a linear layer. \textsc{GroupCover} applies a dense $r\times r$ inverse transformation within each group of output features. Its recovery cost is $O(lnr)$. We use the default group size $r=4$, so each group requires more multiplications inside the TEE.

For linear layers, \textsc{ArrowCloak} \cite{wang_game_2025} computes
\begin{equation}
Y=Y'\Pi^{-1}D_1^{-1}
  -(Xv)\mathbf{1}_n^\top D_2D_1^{-1}.
\end{equation}
Inverse permutation and scaling cost $O(ln)$, while computing $Xv$ costs $O(ld)$. The complete recovery therefore costs $O(l(d+n))$ and includes an input-dependent correction. \textsc{BiShield-TEE} instead computes
\begin{equation}
Y=T_m^{-1}Y'T_r^{-1}
 =D_m^{-1}H_mY'H_rD_r^{-1}.
\end{equation}
Its diagonal scalings and rank-one Householder updates act directly on $Y'$ with $O(ln)$ recovery cost.

The attention-score paths differ as well. For $S=QK^\top$, the obfuscation of $Q$ and $K$ in \textsc{ArrowCloak}~\cite{wang_game_2025} yields
\begin{equation}
S'=\Pi_1(D_1SD_3+\mathrm{OBS})\Pi_2,
\qquad
S=D_1^{-1}
\bigl(\Pi_1^{-1}S'\Pi_2^{-1}-\mathrm{OBS}\bigr)D_3^{-1},
\end{equation}
where
\begin{equation}
\begin{aligned}
\mathrm{OBS}={}&D_2\mathbf{1}_l v_QK^\top D_3
+D_1Qv_K^\top\mathbf{1}_l^\top D_4\\
&+D_2\mathbf{1}_l v_Qv_K^\top\mathbf{1}_l^\top D_4.
\end{aligned}
\end{equation}
Here $\mathbf{1}_l$ is the all-ones vector, $v_Q$ and $v_K$ are masking vectors, $D_i$ are diagonal matrices, and $\Pi_i$ are permutation matrices. The first two correction terms depend on the current $K$ and $Q$, respectively, while the third accounts for the interaction between their masks.

In \textsc{BiShield-TEE}, the coupled transformations yield $S'=T_QST_K^\top$. The TEE recovers the scores directly:
\begin{equation}
S=T_Q^{-1}S'(T_K^\top)^{-1}
 =D_Q^{-1}H_QS'H_KD_K^{-1}.
\end{equation}
This recovery uses diagonal scalings and rank-one Householder updates without constructing $Q$- or $K$-dependent correction terms. The difference in recovery operations is consistent with the end-to-end latency results in Table~\ref{tab:latency-comparison}.

\subsection{Defense and Attack Hyperparameters}
For \textsc{GroupCover} and \textsc{ArrowCloak}, we follow the hyperparameters reported in their original papers; in particular, \textsc{GroupCover} uses the default group size of $4$. For \textsc{BiShield-TEE}, the Householder matrices are randomly generated, and the diagonal entries of the scale matrices are sampled log-uniformly from $[0.1, 10]$ with signs drawn uniformly at random, yielding obfuscation factors that are balanced in magnitude on a multiplicative scale.

\subsection{Implementation}
\label{sec:implement}
All models are trained on a single NVIDIA A800-80G GPU using PyTorch 2.7.1 with CUDA 12.8. For real-device inference evaluation, our implementation builds upon the Gramine framework~\cite{tsai2017graphene}, running on a workstation equipped with an Intel Xeon W-1250 CPU with SGX1 and 93.5 MiB of physical EPC, and an NVIDIA GeForce RTX 5060 Ti GPU.

\section{Sensitivity to Victim--Pretrained Correlation}
\label{app:correlation}

Our reconstruction attacks on \textsc{GroupCover} and \textsc{ArrowCloak} exploit the structural relationship between the victim weights $\mathbf{W}_{\mathrm{vic}}$ and the publicly available pre-trained weights $\mathbf{W}_{\mathrm{pre}}$. We assess how sensitive this exploitation is by varying the magnitude of the fine-tuning increment.

\paragraph{Experimental setup.}
We fine-tune ViT-Base on CIFAR-100 with a scaled fine-tuning update
\begin{equation}
\mathbf{W}_{\mathrm{vic}} = \mathbf{W}_{\mathrm{pre}} + \alpha\cdot\boldsymbol{\Delta},
\end{equation}
where $\boldsymbol{\Delta}$ is the standard fine-tuning update under our experiment configuration, and $\alpha\ge 0$ controls the strength of the deviation from $\mathbf{W}_{\mathrm{pre}}$. We measure the representational similarity between the victim and the pre-trained model using Centered Kernel Alignment (CKA)~\cite{kornblith2019similarity}, where higher CKA indicates greater similarity. We then run the reconstruction attacks of Section~\ref{sec:attacks} at each $\alpha$ and report attack accuracy alongside the victim's own accuracy.

\begin{figure}[htbp]
\centering
\includegraphics[width=\linewidth]{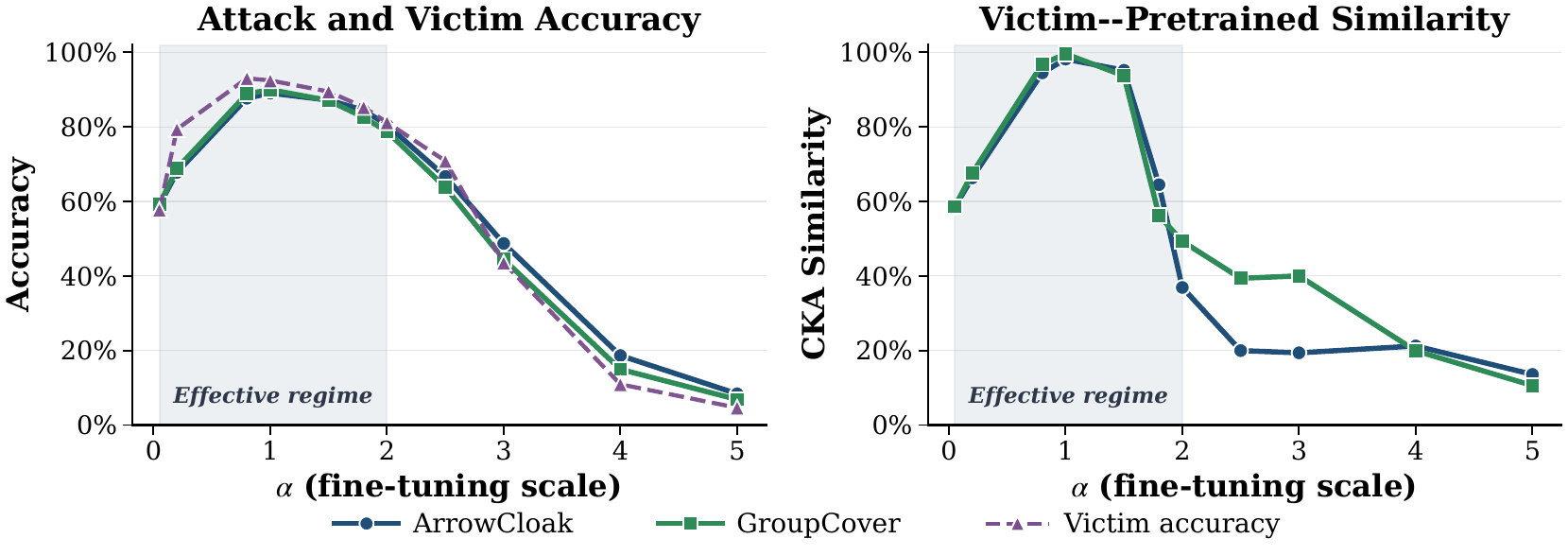}
\caption{Attack and victim accuracy (left), and victim--pre-trained CKA similarity (right), as functions of the fine-tuning scale $\alpha$ on ViT-Base / CIFAR-100. The shaded region marks the effective attack regime ($\alpha\le 2$). Attack accuracy tracks the victim's own accuracy closely across the full range, even when CKA has dropped substantially.}
\label{fig:correlation_ablation}
\end{figure}

\paragraph{Results.}
Figure~\ref{fig:correlation_ablation} reveals three regimes.

\emph{(i) Effective attack regime ($\alpha\le 2$).} Across the entire range relevant to standard fine-tuning, attack accuracy stays high and tracks the victim's own accuracy closely. Notably, this tracking persists even as CKA drops substantially: at $\alpha=1.8$ and $\alpha=2.0$, CKA falls to roughly $0.5$--$0.6$, indicating that $\mathbf{W}_{\mathrm{vic}}$ and $\mathbf{W}_{\mathrm{pre}}$ have already diverged considerably in their internal representations, yet attack accuracy remains within a few percentage points of the victim. The attacks therefore recover the victim's functional behavior even when its structural overlap with $\mathbf{W}_{\mathrm{pre}}$ is far from tight.

\emph{(ii) Boundary regime ($\alpha\approx 2$).} As $\alpha$ approaches $2$, both attack accuracy and CKA begin to drop together. However, this regime also corresponds to noticeable degradation of $\mathbf{W}_{\mathrm{vic}}$ itself, with victim accuracy falling below $82\%$ at $\alpha=2$.

\emph{(iii) Out-of-scope regime ($\alpha>2$).} For $\alpha\ge 2.5$ the victim model collapses, with accuracy dropping below $70\%$ and continuing to fall sharply. Such a model would not be deployed, so the weakening of attacks in this regime is not a meaningful limitation.

\section{Role of Householder Transformations and Diagonal Scaling}
\label{app:transformation-ablation}

We remove diagonal scaling from \textsc{BiShield-TEE} while retaining the bilateral Householder transformations. We apply the unilateral and adaptive bilateral attacks used in the main experiments and report the strongest attacker result for each model and dataset. Rel. Black is attack accuracy divided by black-box accuracy, so values closer to $1.00\times$ indicate stronger protection.

\begin{table}[htbp]
\centering
\caption{Ablation of diagonal scaling. ``H-only'' uses bilateral Householder transformations without scaling. ``Scale+H'' is the complete \textsc{BiShield-TEE} design.}
\label{tab:transformation-ablation}
\small
\renewcommand{\arraystretch}{1.12}
\begin{tabular*}{0.84\linewidth}{@{\extracolsep{\fill}}llccccc@{}}
\toprule
\multirow{2}{*}{\textbf{Model}}
& \multirow{2}{*}{\textbf{Dataset}}
& \textbf{Black-box}
& \multicolumn{2}{c}{\textbf{H-only}}
& \multicolumn{2}{c}{\textbf{Scale+H}} \\
\cmidrule(lr){4-5}\cmidrule(lr){6-7}
& & \textbf{Acc.}
& \textbf{Acc.} & \textbf{Rel. Black}
& \textbf{Acc.} & \textbf{Rel. Black} \\
\midrule
ViT-Base
& CIFAR-10  & 63.4\% & 68.11\% & $1.07\times$ & 63.6\% & $1.00\times$ \\
& CIFAR-100 & 59.7\% & 72.62\% & $1.22\times$ & 59.7\% & $1.00\times$ \\
& Food-101  & 59.2\% & 68.96\% & $1.16\times$ & 59.2\% & $1.00\times$ \\
\midrule
ViT-Large
& CIFAR-100 & 68.8\% & 81.23\% & $1.18\times$ & 68.8\% & $1.00\times$ \\
& Food-101  & 65.3\% & 74.90\% & $1.15\times$ & 65.3\% & $1.00\times$ \\
\midrule
GPT2-Base
& SST-2 & 63.0\% & 83.37\% & $1.32\times$ & 63.0\% & $1.00\times$ \\
& QQP   & 65.6\% & 73.89\% & $1.13\times$ & 65.6\% & $1.00\times$ \\
& QNLI  & 53.7\% & 63.72\% & $1.19\times$ & 53.7\% & $1.00\times$ \\
& MNLI  & 47.7\% & 61.03\% & $1.28\times$ & 47.7\% & $1.00\times$ \\
\midrule
Llama3.2-1B
& GSM8K  & 6.7\%  & 8.57\%  & $1.28\times$ & 7.0\%  & $1.04\times$ \\
& IFEval & 12.1\% & 23.83\% & $1.97\times$ & 12.1\% & $1.00\times$ \\
\midrule
Llama3.2-3B
& GSM8K  & 27.7\% & 31.46\% & $1.14\times$ & 28.0\% & $1.01\times$ \\
& IFEval & 10.5\% & 24.86\% & $2.37\times$ & 10.5\% & $1.00\times$ \\
\midrule
\textbf{Average}
& & -- & -- & $\mathbf{1.34\times}$ & -- & $\mathbf{1.00\times}$ \\
\bottomrule
\end{tabular*}
\end{table}

\paragraph{Ablation Results.}
Householder-only protection allows limited but measurable recovery, with a mean Rel. Black of $1.34\times$ and a maximum of $2.37\times$ on Llama3.2-3B/IFEval. Under the corresponding attacks, the mean ratios for \textsc{GroupCover} and \textsc{ArrowCloak} both exceed $2.4\times$. Bilateral Householder transformations therefore provide substantial protection, but do not consistently reduce attack performance to the black-box level. Adding diagonal scaling reduces the mean ratio to $1.00\times$.

\paragraph{Role of Householder Transformations.}
Bilateral Householder transformations introduce ambiguity into key recovery. However, key ambiguity does not necessarily imply a large difference between the recovered and original weights. A Householder reflection changes the weights along only one direction and preserves their Frobenius norm and singular values. For a uniformly random reflection direction, the expected relative weight displacement decreases as the dimension grows. Thus, in high dimensions, Householder-only protection can obscure the key while leaving the weights relatively close to their original values. This is consistent with the residual recovery gains in Table~\ref{tab:transformation-ablation}.

\paragraph{Role of Diagonal Scaling.}
The full design uses $T_l=H_lD_l$ and $T_r=D_rH_r$. Diagonal scaling changes magnitudes across dimensions and can increase reconstruction errors under an incorrect key. Scaling alone, however, has a separable row-and-column structure. When relative fine-tuning changes are small, element-wise ratios between the protected and pretrained weights approximately reveal the row and column scales. These scales can be estimated up to a reciprocal global factor, which cancels during weight recovery. The two components therefore serve complementary roles: Householder transformations introduce key ambiguity, while diagonal scaling can make incorrect recovery less accurate.

\section{Background on Integrity Verification and One-Time Pad Protection}
\label{app:background}

\subsection{Integrity Check}
\label{app:integrity_check}

Integrity checking \cite{tramer_slalom_2019} aims to ensure that computations outsourced from the TEE to an untrusted component are performed correctly. 
In general, the TEE delegates part of the computation and subsequently verifies the returned results using Freivalds' algorithm. 
Rather than recomputing the full operation, the verifier checks consistency between the input, the output and a small amount of locally generated randomness. 
If the verification fails, the TEE rejects the result and aborts the execution.

Such integrity checks are typically probabilistic: an incorrect result is detected with high probability, while the verification cost remains significantly lower than that of recomputation. 
This approach enables efficient integrity protection for linear operations commonly found in neural network inference, while keeping the trusted computing base small and lightweight.

\subsection{One-Time Pad Protection}
\label{app:otp_protection}

One-time pad (OTP) techniques \cite{tramer_slalom_2019} are commonly used to protect intermediate results during outsourced computation. Given an intermediate value $\mathbf{X} \in \mathbb{R}^{n \times n}$, the TEE prepares a random mask $\mathbf{R} \in \mathbb{R}^{n \times n}$, sends $\tilde{\mathbf{X}} = \mathbf{X} + \mathbf{R}$ to the untrusted component, and recovers the true output as $\mathbf{Y} = \tilde{\mathbf{Y}} - f(\mathbf{R})$, where $f(\cdot)$ is the outsourced operation and $f(\mathbf{R})$ is pre-computed inside the TEE. However, vanilla OTP suffers from two practical issues. First, the mask $\mathbf{R}$ has the same dimensionality as the protected value, so its storage cost scales quadratically with model size and quickly becomes prohibitive on memory-constrained TEEs. Second, refreshing $\mathbf{R}$ frequently is expensive, but reusing the same mask across queries weakens the OTP security guarantee.

To mitigate both issues, low-rank OTP \cite{xiong2025loro} approximates the full mask as the product of two low-rank factors,
\[
\mathbf{R} \approx \mathbf{U}\mathbf{V}^{\top}, \quad \mathbf{U} \in \mathbb{R}^{n \times r}, \; \mathbf{V} \in \mathbb{R}^{n \times r}, \; r \ll n,
\]
reducing storage from $O(n^2)$ to $O(nr)$ and allowing the TEE to refresh or regenerate $\mathbf{U}, \mathbf{V}$ with substantially lower overhead. This enables higher-frequency mask rotation and improves practical security for memory-constrained on-device deployments.

\newpage

\section*{NeurIPS Paper Checklist}

\begin{enumerate}

\item {\bf Claims}
    \item[] Question: Do the main claims made in the abstract and introduction accurately reflect the paper's contributions and scope?
    \item[] Answer: \answerYes{} 
    \item[] Justification: The main contributions have been concluded in the final of Introduction (Section~\ref{sec:intro}).

\item {\bf Limitations}
    \item[] Question: Does the paper discuss the limitations of the work performed by the authors?
    \item[] Answer: \answerYes{} 
    \item[] Justification: Limitations of the work are included in Section~\ref{sec:discussion}.

\item {\bf Theory assumptions and proofs}
    \item[] Question: For each theoretical result, does the paper provide the full set of assumptions and a complete (and correct) proof?
    \item[] Answer: \answerYes{} 
    \item[] Justification: The detailed proofs are deferred to Appendix~\ref{groupcover_proof} and Appendix~\ref{app:security}.

    \item {\bf Experimental result reproducibility}
    \item[] Question: Does the paper fully disclose all the information needed to reproduce the main experimental results of the paper to the extent that it affects the main claims and/or conclusions of the paper (regardless of whether the code and data are provided or not)?
    \item[] Answer: \answerYes{} 
    \item[] Justification: We provide full experimental details in Appendix~\ref{appendix:setup}.

\item {\bf Open access to data and code}
    \item[] Question: Does the paper provide open access to the data and code, with sufficient instructions to faithfully reproduce the main experimental results, as described in supplemental material?
    \item[] Answer: \answerYes{} 
    \item[] Justification: Our data and code are available at \url{https://github.com/laduiw/BiShield-TEE}.

\item {\bf Experimental setting/details}
    \item[] Question: Does the paper specify all the training and test details (e.g., data splits, hyperparameters, how they were chosen, type of optimizer) necessary to understand the results?
    \item[] Answer: \answerYes{} 
    \item[] Justification: All experimental details are provided in Appendix~\ref{appendix:setup}.


\item {\bf Experiment statistical significance}
    \item[] Question: Does the paper report error bars suitably and correctly defined or other appropriate information about the statistical significance of the experiments?
    \item[] Answer: \answerYes{} 
    \item[] Justification: As our primary focus is on reporting model accuracy, error bars may not adequately capture the strongest performance of adversaries. Instead, we report the best results across five independent runs, which provides a more informative assessment.

\item {\bf Experiments compute resources}
    \item[] Question: For each experiment, does the paper provide sufficient information on the computer resources (type of compute workers, memory, time of execution) needed to reproduce the experiments?
    \item[] Answer: \answerYes{} 
    \item[] Justification: Implementation details, including the experimental environment and hardware setup, are described in Appendix~\ref{sec:implement}.
    
\item {\bf Code of ethics}
    \item[] Question: Does the research conducted in the paper conform, in every respect, with the NeurIPS Code of Ethics \url{https://neurips.cc/public/EthicsGuidelines}?
    \item[] Answer: \answerYes{} 
    \item[] Justification: We follow the NeurIPS Code of Ethics.

\item {\bf Broader impacts}
    \item[] Question: Does the paper discuss both potential positive societal impacts and negative societal impacts of the work performed?
    \item[] Answer: \answerNA{} 
    \item[] Justification: Our work focuses on model privacy. Although the attack techniques may have dual-use implications, they are intended to reveal weaknesses in existing defenses and guide the design of more secure systems.
    
\item {\bf Safeguards}
    \item[] Question: Does the paper describe safeguards that have been put in place for responsible release of data or models that have a high risk for misuse (e.g., pre-trained language models, image generators, or scraped datasets)?
    \item[] Answer: \answerNA{} 
    \item[] Justification: We do not release any models or datasets.

\item {\bf Licenses for existing assets}
    \item[] Question: Are the creators or original owners of assets (e.g., code, data, models), used in the paper, properly credited and are the license and terms of use explicitly mentioned and properly respected?
    \item[] Answer: \answerYes{} 
    \item[] Justification: All assets used in this work are publicly available for research purposes, and their licenses can be found on Hugging Face.

\item {\bf New assets}
    \item[] Question: Are new assets introduced in the paper well documented and is the documentation provided alongside the assets?
    \item[] Answer: \answerNA{} 
    \item[] Justification: No new assets are released.

\item {\bf Crowdsourcing and research with human subjects}
    \item[] Question: For crowdsourcing experiments and research with human subjects, does the paper include the full text of instructions given to participants and screenshots, if applicable, as well as details about compensation (if any)? 
    \item[] Answer: \answerNA{} 
    \item[] Justification: No crowdsourcing nor research with human subjects are involved.

\item {\bf Institutional review board (IRB) approvals or equivalent for research with human subjects}
    \item[] Question: Does the paper describe potential risks incurred by study participants, whether such risks were disclosed to the subjects, and whether Institutional Review Board (IRB) approvals (or an equivalent approval/review based on the requirements of your country or institution) were obtained?
    \item[] Answer: \answerNA{} 
    \item[] Justification: No crowdsourcing nor research with human subjects are involved.

\item {\bf Declaration of LLM usage}
    \item[] Question: Does the paper describe the usage of LLMs if it is an important, original, or non-standard component of the core methods in this research? Note that if the LLM is used only for writing, editing, or formatting purposes and does \emph{not} impact the core methodology, scientific rigor, or originality of the research, declaration is not required.
    \item[] Answer: \answerNA{} 
    \item[] Justification: Our workflow and algorithms do not involve the use of LLMs.

\end{enumerate}

\end{document}